%% file: Instanton_paper.tex
\documentclass[11pt]{article}
\usepackage[utf8x]{inputenc}	%for german Umlaute
\usepackage{amssymb}
\usepackage{mathtools}		%for definition ":="
\usepackage{dsfont}		%for math modes
\usepackage{stmaryrd}		%GIT quotient
\usepackage{cite}		%for better citing
\usepackage{hyperref}		%for links on refs etc.
\usepackage{enumerate}
\catcode`@=11
\input{notation}
\allowdisplaybreaks

\begin{document}
\begin{titlepage}
\setcounter{page}{0}
\begin{flushright}
ITP--UH--05/15
\end{flushright}

\vskip 2cm

\begin{center}

{\Large\bf Instantons on Calabi-Yau cones 
}

\vspace{15mm}

{\large Marcus Sperling} 
\\[5mm]
\noindent {\em Institut f\"ur Theoretische Physik} \\
{\em Leibniz Universit\"at Hannover}\\
{\em Appelstra\ss e 2, 30167 Hannover, Germany}\\
Email: {\tt marcus.sperling@itp.uni-hannover.de}
\\[5mm]

\vspace{15mm}

\begin{abstract}
\input{abstract}
\end{abstract}

\end{center}

\end{titlepage}

{\baselineskip=12pt
\tableofcontents
}

%%%%%%%%%%%%%%%%%%%%%%%%%%%%%%%%%%%%%%%%%%%%%%%%%%%%%%%%%%%%%%%%%%%%%%%%%%%%%%%%
  \input{introduction}
%%%%%%%%%%%%%%%%%%%%%%%%%%%%%%%%%%%%%%%%%%%%%%%%%%%%%%%%%%%%%%%%%%%%%%%%%%%%%%%%
  \input{geometry}
%%%%%%%%%%%%%%%%%%%%%%%%%%%%%%%%%%%%%%%%%%%%%%%%%%%%%%%%%%%%%%%%%%%%%%%%%%%%%%%%
  \input{instantons_on_cone}
%%%%%%%%%%%%%%%%%%%%%%%%%%%%%%%%%%%%%%%%%%%%%%%%%%%%%%%%%%%%%%%%%%%%%%%%%%%%%%%%
  \input{remarks}
%%%%%%%%%%%%%%%%%%%%%%%%%%%%%%%%%%%%%%%%%%%%%%%%%%%%%%%%%%%%%%%%%%%%%%%%%%%%%%%%
  \input{discussion}
%%%%%%%%%%%%%%%%%%%%%%%%%%%%%%%%%%%%%%%%%%%%%%%%%%%%%%%%%%%%%%%%%%%%%%%%%%%%%%%%
  \begin{appendix}
  \input{appendix}
  \end{appendix}

 \bibliographystyle{JHEP}     % Zitierstil: alpha = [Nam88]
 {\footnotesize{\bibliography{references}}}

\end{document}

%% file: notation.tex
\renewcommand{\section}{\@startsection{section}{1}{0pt}{\medskipamount}
{\medskipamount}{\large\bf}}
\numberwithin{equation}{section}
\catcode`@=12

\newcommand{\G}{{\rm G}}

\newcommand{\su}{{{\rm SU}(2)}}

\newcommand{\sut}{{{\rm SU}(3)}}
\newcommand{\uo}{{{\rm U}(1)}}

\newcommand{\glrm}{{{\rm GL}}}
\newcommand{\glrmL}{{{\mathfrak{gl}}}}
\newcommand{\urm}{{{\rm U}}}

\newcommand{\surm}{{{\rm SU}}}
\newcommand{\surmL}{{{\mathfrak{su}}}}

\newcommand{\mfrak}{{\mathfrak{m}}}
\newcommand{\gfrak}{{\mathfrak{g}}}

\newcommand{\Hom}{{\rm Hom}}
\newcommand{\End}{{\rm End}}
\newcommand{\diag}{{\rm diag}}
\newcommand{\C}{\mathbb C}
\newcommand{\R}{\mathbb R}
\newcommand{\Z}{\mathbb Z}
\newcommand{\Zcal}{{\cal Z}}
\newcommand{\Acal}{{\cal A}}

\newcommand{\Mcal}{{\cal M}}

\newcommand{\Fcal}{{\cal F}}

\newcommand{\T}{{\cal T}}
\newcommand{\Scal}{{\cal S}}
\newcommand{\Gcal}{{\cal G}}
\newcommand{\Xcal}{{\cal X}}
\newcommand{\Dcal}{{\cal D}}
\newcommand{\Ycal}{{\cal Y}}
\newcommand{\Ucal}{{\cal U}}

\def\im{{\rm i}}
\def\diff{{\rm d}}
\def\Diff{{\rm D}}
\def\tr{{\rm tr}}
\def\and{\quad\textrm{and}\quad}
\def\for{\qquad\textrm{for}\quad}
\def\with{\qquad\textrm{with}\quad}
\newcommand{\Ad}{\mathrm{Ad}}

  \def\dim{\mathrm{dim}}
  \newcommand{\mfd}[1]{M^{#1}}
  \newcommand{\diffZbar}{\diff_{\Zcal}}
  \newcommand{\diffZ}{\bar{\diff}_{\Zcal}}
  \newcommand{\diffYbar}[1]{\diff_{#1}}
  \newcommand{\diffY}[1]{\bar{\diff}_{#1}}  
\newenvironment{Proof}{ \vskip 0.1pt \begin{footnotesize}\noindent 
\textbf{Proof:}}{
\hfill $\Box$ \hfil \end{footnotesize} \vskip 0.1pt
 }  

%% file: abstract.tex
\noindent
The Hermitian Yang-Mills equations on certain vector bundles over 
Calabi-Yau cones can be reduced to a set of matrix equations; in fact, these 
are Nahm-type equations. The latter can be analysed further by generalising 
arguments of Donaldson and Kronheimer used in the study of the original 
Nahm equations. Starting from certain equivariant connections, we show that the 
full set of instanton equations reduce, with a unique gauge transformation, to 
the holomorphicity condition alone.

%% file: introduction.tex
\section{Introduction}
Instantons have proven to be interesting both for mathematicians and 
physicists. Starting from the seminal work~\cite{Donaldson:1983} by Donaldson, 
anti-self-dual Yang-Mills connections provided a new topological invariant for 
four-manifolds. However, the moduli spaces of higher-dimensional instantons are 
still not fully understood.

From a physics perspective, instantons describe non-perturbative Yang-Mills 
configurations in 
various settings\cite{Rajaraman:1982is,Manton:2004tk,Weinberg:2012}. Focusing, 
for example, on heterotic string theory and compactifications thereof, the 
notion of instantons appears naturally in the so-called BPS-equations. In the 
simplest case, it is necessary to specify a $6$-dimensional Calabi-Yau manifold 
as well as a Hermitian Yang-Mills (HYM) instanton on a gauge bundle 
over that manifold~\cite{Candelas:1985en}. However, due to the appearance of 
phenomenologically problematic moduli, it is physically desirable to relax the 
strict Calabi-Yau condition to one of more general $\sut$-manifolds 
($\sut$-structures with intrinsic torsion), for which the instanton notion 
needs to be adjusted. Examples of those are nearly Kähler and half-flat 
manifolds. For further details on so-called flux compactifications see for 
instance~\cite{Grana:2005jc,Blumenhagen:2006}.

Recently, the study of Sasaki-Einstein 
manifolds~\cite{Gauntlett:2004yd,Gauntlett:2004hh,Cvetic:2005ft,Cvetic:2005vk,
Lu:2005sn} has led to infinitely many explicit metrics on (non-compact) 
Calabi-Yau cones. Since there are no explicit Ricci-flat metrics known on 
compact Calabi-Yau manifolds, metric cones over Sasaki-Einstein spaces provide a 
testing ground for Calabi-Yau compactifications.

Previously, instantons have been discussed on certain cone constructions 
starting from a $\G$-manifold, i.e. a manifold that admits a 
$\G$-structure~\cite{Harland:2009yu,Harland:2010ix,Bauer:2010fia,Haupt:2011mg,
Gemmer:2011cp,Harland:2011zs,Ivanova:2012vz,Bunk:2014kva,Bunk:2014coa}. There 
on the instanton equations have been reduced to a set of matrix 
equations.\footnote{These matrix equations were first introduced 
in~\cite{Ivanova:2012vz} as a generalisation of the results 
in~\cite{Harland:2011zs}. We will refer to these equations as 
\emph{Harland-Ivanova-Nölle-Popov} (HINP) matrix equations.} The aim of 
this paper is to discuss the resulting matrix equations on Calabi-Yau cones 
over a generic Sasaki-Einstein manifold $\mfd{2n+1}$, which carries an 
$\surm(n)$-structure. In particular, the HINP matrix equations conceptually 
comprise three types of equations: (i) the so-called equivariance condition, 
(ii) the holomorphicity condition, and (iii) a stability-like condition. 
Starting 
form solutions to (i), i.e. decomposing the matrices into irreducible 
representations of $\surmL(n)$, we show that it suffices to solve 
(ii) for certain boundary conditions, because (iii) then follows 
by a unique gauge transformation. The arguments presented are a generalisation 
of~\cite{Donaldson:1984,Kronheimer:1989,Kronheimer:1990}.

Instantons on Calabi-Yau cones and their resolutions have 
also been studied in~\cite{Correia:2009,Correia:2010} and, for the particular 
orbifolds $\C^n/\Z_n$, in~\cite{Nibbelink:2007}. However, their setting and 
ansatz are different: on the one hand, \cite{Correia:2009,Correia:2010} 
considered instantons on the \emph{tangent bundle} of a $(2n{+}2)$-dimensional 
Calabi-Yau cone whose structure was largely determined by the $2n$-dimensional 
Einstein-Kähler manifold underlying the Sasaki-Einstein manifold in between. The 
ansatz for the connection was adapted to the isometry of the Calabi-Yau cone, 
and the ``seed'' was the spin connection in the Einstein-Kähler space, which is 
an instanton. On the other hand, certain gauge backgrounds for heterotic 
compactifications  were constructed in~\cite{Nibbelink:2007} by extending 
a flat connection on $\C P^{n-1}$ to $\uo$ and $\urm(n{-}1)$-valued instanton 
connections on the orbifolds. In contrast, the approach 
of~\cite{Ivanova:2012vz}, which is further discussed here, can conceptually 
take \emph{any} instanton on the Sasaki-Einstein manifold as a starting point, 
and the bundle does not need to be the tangent bundle anymore.

This paper is organised as follows: the relevant details on Sasaki-Einstein 
manifolds and the Calabi-Yau cone over it are briefly summarised in 
Section~\ref{sec:Preliminaries}. In addition, the notion of Hermitian 
Yang-Mills instantons is recalled. The main body is 
Section~\ref{sec:Instantons} where we firstly recapitulate the ansatz for the 
connection that reduces the HYM equations to matrix equations. The 
subsequent paragraphs consider the geometry, symmetries and solutions to these 
equations. Section~\ref{sec:conclusion} concludes.

%% file: geometry.tex
\section{Preliminaries}
\label{sec:Preliminaries}
\subsection{Sasaki-Einstein manifolds}
\emph{Sasakian geometry} can be understood as odd-dimensional analogue of 
Kähler geometry; in particular, an odd-dimensional manifold $\mfd{2n+1}$ with a 
Sasakian structure is naturally sandwiched between two different types of 
Kähler geometry in the neighbouring dimensions $2n$ and $2n{+}2$.

Following~\cite{Boyer:2008}, a Sasakian manifold $\mfd{2n+1}$ 
carries a Sasakian structure comprised of the quadruplet $\Scal= (\xi, \eta, 
\Phi, g)$, wherein $\xi$ is the Reeb vector field, $\eta$ the dual 
contact form, $\Phi \in \End(T \mfd{2n+1})$ a tensor, and $g$ a Riemannian 
metric. The defining property for $(\mfd{2n+1},\Scal)$ to be Sasakian is that 
the metric cone $(C(\mfd{2n+1}),\widehat{g}) = (\R^+ \times \mfd{2n+1},\diff r^2 
+ r^2 g)$ is Kähler, i.e. the holonomy group of the Levi-Civita connection on 
the cone is $\urm(n{+}1)$. The (compatible) complex structure $J_c$ on the cone 
acts via $J_c (r \partial_r) = \xi$ and $J_c(X)= \Phi(X) -\eta(X) 
r\partial_r$ for any vector field $X$ on $\mfd{2n+1}$.  The corresponding 
Kähler $2$-form is $\tfrac{1}{2} \diff (r^2 \eta)$.

Moreover, considering the contact subbundle $\Dcal = \mathrm{ker}(\eta)\subset 
T\mfd{2n+1}$ one has a complex structure defined by restriction $J_t = \Phi 
|_\Dcal$ and a symplectic structure $\diff \eta$. Hence, $(\Dcal, J_t,\diff 
\eta)$ defines the transverse Kähler structure~\cite{Boyer:2008}.

A \emph{Sasaki-Einstein manifold} is Sasakian and Einstein simultaneously; 
thus, the 
defining property is that the metric cone is Calabi-Yau, i.e. the holonomy 
group on the cone is reduced to $\surm(n{+}1)$.

For the purposes of this paper, it is convenient to understand a 
Sasaki-Einstein manifold $\mfd{2n+1}$ in terms of an 
$\surm(n)$-structure. For this, one has the $1$-form $\eta$ and the $2$-form 
$\omega$, which are related via $\diff \eta = -2\omega$. One can always choose 
a co-frame $\{e^\mu\}=(e^a,e^{2n+1}) $, with $\mu=1,2,\ldots, 2n+1$ and $a=1,2, 
\ldots,2n$, such that these forms are locally given by
\begin{equation}
 \eta = e^{2n+1} \and \omega= e^1 \wedge e^2 +  e^3 \wedge e^4 + \ldots +  
e^{2n-1} \wedge e^{2n} \equiv \frac{1}{2} \omega_{ab} e^{ab} \; 
\end{equation}
and that the metric is 
\begin{equation}
 g= \delta_{\mu \nu} e^\mu \otimes e^\nu = \delta_{ab}  e^a \otimes e^b + 
\eta \otimes \eta \; .
\end{equation}
Moreover, there exists a canonical connection 
$\Gamma^P$ on $T \mfd{2n+1}$ which is metric compatible, is an instanton with 
respect to the $\surm(n)$-structure, and has non-vanishing 
torsion.~\footnote{The torsion components can be related to the components of 
the $3$-form 
$P=\eta\wedge \omega$; hence, the name $\Gamma^P$. However, the torsion is not 
completely antisymmetric itself.} The torsion 
components are given by~\cite{Harland:2011zs}
\begin{equation}
 T^{2n+1}_{a\,b} = -2 \omega_{ab} \and T_{a \, 2n+1}^b = \frac{n+1}{n} 
\omega_{ab} \; .\label{eqn:torsion}
\end{equation}
\subsection{Calabi-Yau metric cone}
First of all, recall the basic properties of a Calabi-Yau manifold 
$\mfd{2n+2}$: as a Calabi-Yau 
space is Kähler, one has the Kähler form, which is an exact 
$(1,1)$-form on $\mfd{2n+2}$. In addition, the Calabi-Yau condition enforces 
the canonical bundle to be trivial, i.e. $K_{\mfd{2n+2}}= \Lambda^{(n+1,0)} 
T^* \mfd{2n+2} \cong \C \times \mfd{2n+2}$. Thus, there exists 
a nowhere vanishing section in $K_{\mfd{2n+2}}$ which translates into a 
$(n{+}1,0)$-form on $\mfd{2n+2}$.

The metric on the metric cone $(C(\mfd{2n+1}),\widehat{g})$ is defined as
\begin{equation}
 \widehat{g}= \diff r^2 + r^2 g = e^{2 t } \left( \diff t^2 + 
\delta_{\mu \nu} e^\mu \otimes e^\nu  \right) = e^{2 t} \widetilde{g}  \; ,
\end{equation}
where the last equality employs a conformal rescaling $r = e^t$ from the 
metric cone with cone coordinate $r \in \R^+$ to the cylinder 
$(\mathrm{Cyl}(\mfd{2n+1}),\widetilde{g})= (\R\times \mfd{2n+1} , \diff 
t^2 + g)$ with coordinate $t \in \R$. Also, we identify $\diff t = 
e^{2n+2}$ and extend the index range $\hat{\mu} = 1,2, \ldots, 2n+1,2n+2 $.  
The Kähler form $ \widehat{\omega}$ on the cone is 
\begin{equation}
 \widehat{\omega} = r^2 \omega + r \eta \wedge \diff r = e^{2 t } \left( 
\omega + \eta \wedge \diff t \right)  = e^{2 t } \widetilde{\omega} \; ,
\end{equation}
which is again related to the Kähler form $\widetilde{\omega}$ on the cylinder.
Next, we introduce a complexified basis on the cotangent bundle of 
$\mathrm{Cyl}(\mfd{2n+1})$ as follows
\begin{equation}
 \theta^{j} = \im e^{2j-1} + e^{2j} \and  \bar{\theta}^{j} = -\im e^{2j-1} +
e^{2j} \for j= 1,2,\ldots , n+1 \; , \label{eqn:def_cplx_forms}
\end{equation}
such that the metric and Kähler form read
\begin{equation}
 \widetilde{g} = \frac{1}{2} \sum_{j = 1}^{n+1} \left( \theta^j \otimes 
\bar{\theta}^j + \bar{\theta}^j \otimes \theta^j  \right) \and 
\widetilde{\omega} = -\frac{\im}{2}\sum_{j = 1}^{n+1}\theta^j \wedge 
\bar{\theta}^j \; .
\end{equation}
The compatible complex structure $J$ acts via $J \theta^j = \im \theta^j$ and 
$J \bar{\theta}^j = - \im \bar{\theta}^j$, such that the compatibility 
relation is $\widetilde{\omega}(\cdot,\cdot) = \widetilde{g}(\cdot , J \cdot)$.

Let us compare the choice~\eqref{eqn:def_cplx_forms} with the 
``canonical choice'' $\theta_{\mathrm{can}}^{j} = e^{2j-1} +\im  e^{2j}$ and 
the canonical complex structure $J_{\mathrm{can}}\theta_{\mathrm{can}}^{j} = 
\im \theta_{\mathrm{can}}^{j}$. The conventions used here correspond to $J = 
-J_{\mathrm{can}}$ such that the $(1,0)$ and $(0,1)$-forms are interchanged, 
which implies that $\widetilde{\omega}(\cdot,\cdot) 
=\widetilde{g}(J_{\mathrm{can}} \cdot , \cdot) = - \widetilde{g}( \cdot 
,J_{\mathrm{can}} \cdot) = \widetilde{g}(\cdot , J \cdot)$ is consistent with 
the above. The reasons for this choice are that we desire a resemblance to the 
treatment of~\cite{Donaldson:1984,Kronheimer:1989,Kronheimer:1990}, while at the 
same time we treat $\diff t$ as the $(2n{+}2)$th basis $1$-form instead of the 
$0$th.
\subsection{Hermitian Yang-Mills instantons}
\label{subsec:HYM_intro}
For the later analysis, the geometric properties of the space of connections 
and the HYM instanton moduli space over a Kähler manifold are recalled. This 
brief account is inspired from~\cite{Atiyah:1982,Deser:2014}. 

\paragraph{Space of connections}
Let $\mfd{2n}$ be a (closed) Kähler manifold of $\dim_\C(M)=n $ and $\Gcal$ a 
compact matrix Lie group. Let $P(\mfd{2n},\Gcal)$ be a 
$\Gcal$-principal bundle over $\mfd{2n}$, $\Acal$ a connection 1-form and 
$\Fcal_{\Acal}= \diff \Acal + \Acal \wedge \Acal$ 
the curvature. 

Let $\mathrm{Int}(P)\coloneqq P \times_\Gcal \Gcal $ be the group bundle (where 
$\Gcal$ 
acts via the internal automorphism $h \mapsto g h g^{-1}$), let 
$\Ad(P)\coloneqq P \times_{\Gcal}\gfrak$ be the Lie algebra bundle 
(where $\Gcal$ acts on $\gfrak$ via the adjoint action), and $E\coloneqq P 
\times_{\Gcal}F$ be an associated vector bundle (where the vector space $F$, 
the typical fibre, carries a $\Gcal$-representation).

Denote the space of all connections on $P$ by $\mathbb{A}(P)$ and note that all 
associated bundles $E$ inherit their space of connections $\mathbb{A}(E)$ from 
$P$. On $\mathbb{A}(P)$ there is a natural action of the gauge group 
$\widehat{\Gcal}$, i.e. the set of automorphisms on $P$ which are trivial on 
the base. With 
\begin{equation}
 \widehat{\Gcal} = \Gamma(\mfd{2n},\mathrm{Int}(P))  
\end{equation}
one has an identification with the space of global sections of the group 
bundle. The action is realised via
\begin{equation}
 \Acal \to \Acal^g = \mathrm{Ad}(g^{-1}) \Acal + g^{-1} \diff g  \for g \in 
\Gamma(\mfd{2n},\mathrm{Int}(P)) \; . \label{eqn:trafo_connection}
\end{equation}
The Lie algebra of the gauge group is then given as 
\begin{equation}
 \widehat{\gfrak} = \Gamma(\mfd{2n},\mathrm{Ad}(P)) \; ,
\end{equation}
and the infinitesimal gauge transformations are given by
\begin{equation}
 \Acal \mapsto \delta \Acal = \diff_\Acal \chi \coloneqq \diff \chi + \left[ 
\Acal, \chi \right] \for \chi \in \Gamma(\mfd{2n},\mathrm{Ad}(P)) \; . 
\label{eqn:inf_gauge_transf}
\end{equation}

Since $\mathbb{A}(P)$ is an affine space, the tangent space $T_{\Acal} 
\mathbb{A}$ for any $\Acal \in \mathbb{A}(P)$ is canonically identified with 
$\Omega^1(\mfd{2n},\mathrm{Ad}(P))$. Further, assuming $\Gcal \hookrightarrow 
\urm(N)$ for some $N\in \mathbb{N}$, implies that the trace is an 
$\Ad$-invariant inner product. 
Hence, a metric on $\mathbb{A}(P)$ is defined via 
\begin{equation}
 \boldsymbol{g}_{|\Acal}(X_1 , X_2 ) \coloneqq \int_{\mfd{2n}} \tr \left( X_1 
\wedge 
\star X_2 \right) \for X_1,X_2 \in T_\Acal \mathbb{A} \; ,
\label{eqn:metric_connections}
\end{equation}
with $\star$ the Hodge-dual on $\mfd{2n}$.
Moreover, the space $\mathbb{A}(P)$ allows for a symplectic structure 
\begin{subequations}
 \label{eqn:symplectic_form_connections}
\begin{equation}
 \boldsymbol{\omega}_{|\Acal} (X_1,X_2) \coloneqq \int_{\mfd{2n}} \tr \left( 
X_1 \wedge 
X_2 \right) \wedge \frac{ \omega^{n-1}}{(n-1)!}  \for X_1,X_2 \in T_\Acal 
\mathbb{A}\; . 
\end{equation}
Since $\boldsymbol{\omega}$ is completely base-point independent (on 
$\mathbb{A}$), $\boldsymbol{\omega}$ is in fact a 
symplectic form. In addition, one can check that $X\wedge \frac{ 
\omega^{n-1}}{(n-1)!}  = \star J(X)$ holds for any $ X \in T_\Acal 
\mathbb{A} $, where $J$, the (canonical) complex structure of $\mfd{2n}$, acts 
only the $1$-form part of $X$. This allows to reformulate the symplectic 
structure as 
\begin{equation}
 \boldsymbol{\omega}_{|\Acal} (X_1,X_2) = \int_{\mfd{2n}} \tr \left( 
X_1 \wedge \star J (X_2) \right) \for X_1,X_2 \in T_\Acal 
\mathbb{A}\; .
\end{equation}
\end{subequations}
Moreover, it implies that $\boldsymbol{\omega}$ is non-degenerate as 
$\boldsymbol{\omega}_{|\Acal} (X_1,X_2) = \boldsymbol{g}_{|\Acal}(X_1 , J(X_2) 
)$ holds for any $X_1,X_2$ and any $\Acal$. Consequently, 
$(\mathbb{A},\boldsymbol{g},\boldsymbol{\omega})$ is an infinite-dimensional 
Riemannian, symplectic manifold, which is equipped with compatible 
$\widehat{\Gcal}$-action.
\paragraph{Holomorphic structure}
Next, consider the restriction to connections on $E \xrightarrow{\simeq F} 
M$ which satisfy the so-called holomorphicity condition
\begin{equation}
 \Fcal_\Acal^{2,0}=0 \and \Fcal_\Acal^{0,2}=0 \; . \label{eqn:holo_cond}
\end{equation}
It is well-known that this condition is equivalent to the existence of a 
holomorphic structure on $E$, i.e. a Cauchy-Riemann operator 
$\bar{\partial}_E := \bar{\partial} + A^{0,1}$ that satisfies the Leibniz-rule 
as well as $\bar{\partial}_E \circ \bar{\partial}_E =0$. Thus, having a 
$\Gcal$-bundle with a holomorphic connection induces a holomorphic 
$\Gcal^\C$-bundle. If $\mfd{2n}$ is also Calabi-Yau, then the
condition~\eqref{eqn:holo_cond} is equivalent to $\Omega \wedge \Fcal_\Acal 
=0$, where $\Omega$ is a holomorphic $(n,0)$-form.

Define the subspace of holomorphic connections as
\begin{equation}
 \mathbb{A}^{1,1} = \left\{ \Acal \in \mathbb{A}(E) : 
\Fcal_\Acal^{0,2} = - \left( \Fcal_\Acal^{2,0} \right)^\dagger =0  
\right\} \subset \mathbb{A}(E) \; .
\end{equation}
This definition employs the underlying complex structure on 
$\mfd{2n}$. Moreover, one can show that $ \mathbb{A}^{1,1}$ is an 
infinite-dimensional Kähler space, i.e. $ \boldsymbol{g}$ is a Hermitian 
metric and the symplectic form $\boldsymbol{\omega}$ is Kähler. We note that 
these objects descend from $\mathbb{A}$ to $\mathbb{A}^{1,1}$ simply by 
restriction. The compatible 
complex structure $\boldsymbol{J}$ (with $\boldsymbol{\omega}(\cdot,\cdot) = 
\boldsymbol{g}(\boldsymbol{J}\cdot,\cdot)$) can be read off 
from~\eqref{eqn:metric_connections} and~\eqref{eqn:symplectic_form_connections} 
to be 
\begin{equation}
 \boldsymbol{J}_{|\Acal}(X) = -J(X) \for X\in T_\Acal 
\mathbb{A} \; , \label{eqn:cplx_str_connections}
\end{equation}
i.e. it is base point independent.
\paragraph{Moment map}
The space $\mathbb{A}^{1,1}$ inherits the $\widehat{\Gcal}$-action from 
$\mathbb{A}$ and since it has a symplectic form, i.e. the Kähler form, one can 
introduce a moment map
\begin{equation}%
 \begin{split}%
 \mu : \mathbb{A}^{1,1}  &\to  \widehat{\gfrak}^* 
\cong \Omega^{2n}(\mfd{2n},\mathrm{Ad}(P)) \\
 \Acal &\mapsto \mathcal{F}_\Acal \wedge \frac{ \omega^{n-1}}{(n-1)!} \; .
 \end{split}   \label{eqn:moment_map_aux}%
\end{equation}%
We see that $\mu$ is $\widehat{\Gcal}$-equivariant by construction. 
Nonetheless, for this to be a moment map of the $\widehat{\Gcal}$-action, one 
needs to verify the defining property
\begin{equation}
 (\phi,\Diff \mu_{| \Acal})(\psi) = \iota_{\phi^\natural}  
\boldsymbol{\omega}_{|\Acal} (\psi) \; , \label{eqn:aux_moment_map}
\end{equation}
where $\phi \in \Gamma(\mfd{2n},\mathrm{Ad}(P))$ an element of the gauge Lie 
algebra,
$\phi^\natural$ be the corresponding vector field on $ \mathbb{A}^{1,1}$ and 
$\psi \in \Omega^1(\mfd{2n},\mathrm{Ad}(P))$ a tangent vector at the base point 
$\Acal$. Moreover, the duality pairing $(\cdot, \cdot)$ of $\widehat{\gfrak}$ 
and its dual is defined via the integral over $\mfd{2n}$ and the invariant 
product on $\gfrak$. Generalising the arguments from~\cite{Atiyah:1982}, one can 
prove that $\mu$ is indeed a moment map for the $\widehat{\Gcal}$-action on 
$\mathbb{A}^{1,1}$. 
Firstly, in the definition of $\mu$ only $\Fcal_\Acal$ is base point dependent, 
and a 
standard computation gives $\Fcal_{\Acal + t \, \psi} = \Fcal_\Acal + t\, 
\diff_\Acal \psi + 
\tfrac{1}{2}\, t^2 \, \psi\wedge\psi$ so that $\Diff \Fcal_{|\Acal}(\psi) = 
\big( \frac{\diff}{\diff t} \Fcal_{\Acal + t \, \psi} \big)_{|{t=0}} 
=\diff_\Acal \psi $. Thus the left-hand side of~\eqref{eqn:aux_moment_map} is  
$(\phi,\Diff \mu_{| \Acal})(\psi)= \int_M \, \tr \big( (\diff_\Acal 
\psi) \wedge \phi \big) \wedge \frac{ \omega^{n-1}}{(n-1)!} $. Secondly, the 
vector field $\phi^\natural$ can be read off 
from~\eqref{eqn:inf_gauge_transf} to be $\phi^\natural_{|\Acal} = \diff_\Acal 
\phi \in \Omega^1(M , \mathrm{Ad}(P))$. Hence the right-hand side is
$\iota_{\phi^\natural}  
\boldsymbol{\omega}_{|\Acal} (\psi)= \int_M\, \tr \big( (\diff_\Acal \phi) 
\wedge \psi \big) \wedge \frac{\omega^{n-1}}{(n-1)!} $. But from $\int_M \, 
\diff 
\left( \tr \left( \psi \wedge \phi \right) \wedge \frac{\omega^{n-1}}{(n-1)!} 
\right) =0$ by Stokes' theorem\footnote{For the non-compact Calabi-Yau cone of 
this paper, the boundary term arising by Stokes' theorem will be cancelled be 
restriction to \emph{framed} gauge transformations. See 
Section~\ref{subsec:Geo_structure}.}
and $\diff \omega =0$ one has $\int_M\, \tr \big( (\diff_\Acal \psi) 
\wedge \phi\big) \wedge \frac{\omega^{n-1}}{(n-1)!} = - \int_M \tr\, \big(  
\psi \wedge (\diff_\Acal \phi )\big) \wedge \frac{\omega^{n-1}}{(n-1)!} $, and 
therefore the
relation~\eqref{eqn:aux_moment_map} holds, i.e. $\mu$ is a moment map of the 
$\widehat{\Gcal}$-action on $\mathbb{A}^{1,1}$.

However, one can equally well use the dual map defined by 
\begin{equation}%
 \begin{split}%
 \mu^* : \mathbb{A}^{1,1}  &\to  \widehat{\gfrak} = 
\Omega^0(\mfd{2n},\mathrm{Ad}(P)) 
\\
 \Acal &\mapsto \omega\lrcorner \mathcal{F}_\Acal \; ,
 \end{split} %
  \label{eqn:moment_map_generic}%
\end{equation}%
 which is equivalent to $\mu$ of~\eqref{eqn:moment_map_aux} due to
\begin{equation}
 \Fcal_\Acal \wedge \omega^{n-1}  = \frac{1}{n} (\omega \lrcorner 
\Fcal_\Acal) \omega^n \; .
\end{equation}
Thus, we will no longer explicitly distinguish between $\mu$ and $\mu^*$.

For $\Xi \in \mathrm{Centre}( \widehat{\mathfrak{g}})$, we know $\mu^{-1}(\Xi) 
\subset \mathbb{A}^{1,1} $ defines a sub-manifold which allows for a 
$\widehat{\Gcal}$-action. The quotient 
\begin{equation}
\mathbb{A}^{1,1}\sslash  \widehat{\Gcal}  \equiv \mu^{-1}(\Xi) \slash
\widehat{\Gcal} 
\end{equation}
is well-defined and, moreover, is a Kähler manifold, as the Kähler form and the
complex structure descend from $\mathbb{A}^{1,1} $. 

We recognise the zero-level set as the \emph{Hermitian Yang-Mills} moduli space. 
In other words, the HYM equations consist of the holomorphicity 
conditions~\eqref{eqn:holo_cond} together with the so-called stability condition
 \begin{equation}
 \mu(\Fcal_\Acal) = \Fcal_\Acal \wedge \frac{\omega^{n-1}}{(n-1)!} = 0 = 
\mu^*(\Fcal_\Acal)= \omega \lrcorner \Fcal_\Acal\; . 
\label{eqn:stab_cond}
\end{equation}
By well-known theorems~\cite{Donaldson:1985,Uhlenbeck:1986,Donaldson:1987}, a 
holomorphic vector bundle admits a solution to the HYM equations if and 
only if these bundles are (poly-)stable in the algebraic geometry sense.
\paragraph{Complex group action}
 As the $\widehat{\Gcal}$-action on $\mathbb{A}^{1,1}$ preserves the Kähler 
structure, one can extend to an $\widehat{\Gcal}^\C$-action on 
$\mathbb{A}^{1,1}$. In other words, the holomorphicity conditions 
$\Fcal_\Acal^{0,2}=0$ are invariant under the action 
of the complex gauge group
\begin{equation}
 \widehat{\Gcal}^\C =\widehat{\Gcal} \otimes \C \; .
\end{equation}
Let $\Acal \in  \mathbb{A}^{1,1}$, then the orbit $\widehat{\Gcal}_\Acal^\C$ 
of the $\widehat{\Gcal}^\C$-action is 
\begin{equation}
 \widehat{\Gcal}_\Acal^\C = \left\{ \Acal' \in \mathbb{A}^{1,1} \, 
\big| \,  \exists q \in \widehat{\Gcal}^\C : \Acal'=  \Acal^q \right\}.
\end{equation}
A point $\Acal \in  \mathbb{A}^{1,1}$ is called \emph{stable} if 
$\widehat{\Gcal}_\Acal^\C  \cap \mu^{-1}(\Xi) \neq \emptyset$, and we denote 
by $\mathbb{A}^{1,1}_{st}(\Xi) \subset\mathbb{A}^{1,1}$ the set of all 
stable points (for a given $\Xi$). Then, a well-known result (see for 
example~\cite{Thomas:2006}) is
\begin{equation}
  \mathbb{A}^{1,1} \sslash  \widehat{\Gcal}  \equiv \mu^{-1}(\Xi) 
\slash \widehat{\Gcal}  \cong  \mathbb{A}^{1,1}_{st}(\Xi) \slash 
\widehat{\Gcal}^\C \; .
\end{equation}
\paragraph{Remark}
A peculiarity arises for holomorphic bundles $E$ over a compact Kähler manifold 
 $\mfd{2n}$ with non-empty boundary~\cite{Donaldson:1992}. Due to the 
prescription of boundary conditions, the stability condition is automatically 
satisfied for a unitary connection whose curvature is of type $(1,1)$. Hence, 
all points in $\mathbb{A}^{1,1}$ are stable in this case.

In the following we will consider the HYM equations~\eqref{eqn:holo_cond} 
and~\eqref{eqn:stab_cond} on the non-compact Calabi-Yau cones. For these, the 
holomorphicity conditions still imply the existence of a holomorphic structure; 
while the notion of stability is not applicable anymore. Nonetheless, we will 
continue referring to $\omega \lrcorner \Fcal_\Acal =0$ as 
\emph{stability-like} condition.

%% file: instantons_on_cone.tex
\section{Equivariant instantons}
\label{sec:Instantons}
The main focus of this paper lies on the description of the instantons on 
certain vector bundles $E$. However, instead of generic connections the set-up 
will 
be restricted to connections that arise from an instanton on the 
Sasaki-Einstein space $\mfd{2n+1}$ by an extension $X\in 
\Omega^1(\mathrm{Cyl}(\mfd{2n+1});\End(E))$. This extension has to satisfy a 
certain invariance condition. 

The arguments presented in what follows are a generalisation 
of~\cite{Donaldson:1984,Kronheimer:1989,Kronheimer:1990}: i.e. we 
generalise from spherically symmetric instantons on vector bundles over $C(S^3) 
\cong \R^4\backslash \{0\}$ with an $\su$-structure to 
$\surm(n{+}1)$-equivariant instantons on vector bundles over $C(\mfd{2n+1})$ 
with 
an $\surm(n{+}1)$-structure, where $\mfd{2n+1}$ is an arbitrary Sasaki-Einstein 
manifold. Analogous to Donaldson and Kronheimer, it will be necessary to 
consider boundary conditions for the components of the connection 1-form, i.e. 
for the Yang-Mills fields.
\subsection{Ansatz}
\label{subsec:Ansatz}
Let us recall the ansatz presented in~\cite{Ivanova:2012vz} and explicitly 
discussed in~\cite{Bunk:2014coa}. Start from any Sasaki-Einstein manifold 
$\mfd{2n+1}$, i.e. the manifold carries an $\surm(n)$-structure together with a 
canonical connection $\Gamma^P$ on the tangent bundle. The metric cone is 
Calabi-Yau with holonomy $\surm(n{+}1)$, i.e. an integrable 
$\surm(n{+}1)$-structure. By conformal equivalence one can consider 
$\mathrm{Cyl}(\mfd{2n+1})$.

 Consider a complex vector bundle $E \to \mathrm{Cyl}(\mfd{2n+1})$ of 
rank $p$ which has structure group $\surm(n{+}1)$; in particular, that is a 
Hermitian vector bundle where $\Fcal^\dagger = -\Fcal$ and $\tr(\Fcal)=0$ hold 
for the curvature $\Fcal$ of a compatible connection. (In the compact case, one 
would have a vanishing first Chern 
class.) For example, the (holomorphic) tangent 
bundle of the Calabi-Yau cone is such a bundle, but one does not have to 
restrict to this case.
 
 We recall that the connection $1$-forms are $\surmL(n{+}1)$-valued 
$1$-forms on $\mathrm{Cyl}(\mfd{2n+1})$ for any connection $\Acal$ on $E$. The 
ansatz for a connection is
\begin{subequations}
\label{eqn:ansatz_generic}
\begin{equation}
 \Acal = \widehat{\Gamma}^P + X
\end{equation}
where $\widehat{\Gamma}^P$ is the lifted $\surmL(n)$-valued connection on $E$ 
obtained from $\Gamma^P$, i.e. one essentially has to change the representation 
on the fibres. Moreover, on a patch $\Ucal \subset \mathrm{Cyl}(\mfd{2n+1})$ 
with the co-frame $\{ e^{\hat{\mu}}\}$ we employ the local description
\begin{equation}
  X_{|\Ucal}= X_\mu \otimes e^\mu + X_{2n+2} \otimes e^{2n+2} \; ,
\end{equation}
\end{subequations}
where ${X_{\hat{\mu}}}_{|x} \in \End(\C^p)$ for $x\in \Ucal$. Usually 
$X_{2n+2}$ is eliminated by a suitable gauge transformation, but there is no 
harm in not doing so.

The ansatz~\eqref{eqn:ansatz_generic} is a generic connection in the sense that 
the $X_{\hat{\mu}}$ are base-point dependent, skew-Hermitian, traceless 
matrices with nontrivial transformation behaviour under change of 
trivialisation. Hence, \emph{any} connection $\Acal$ on $E$ can be reached 
starting from $\widehat{\Gamma}^P$. 

 Since $\surm(n)$ is a closed subgroup of $\surm(n+1)$, one can choose an 
$\surm(n)$-invariant decomposition
\begin{align}
 \surmL(n+1) = \surmL(n) \oplus \mfrak \with  
\begin{split}
 \surmL(n+1) &= \mathrm{span}\left\{ I_A  \, \big| \, A = 1, \ldots, (n+1)^2 
-1\right\} \; , \\
 \surmL(n) &= \mathrm{span}\left\{ I_\alpha \, \big| \, \alpha= 2n+2, \ldots , 
(n+1)^2\right\} \; ,\\
 \mfrak &= \mathrm{span}\left\{ I_\mu \, \big| \, \mu= 1, \ldots , 2n+1\right\} 
\; ,\end{split}
\end{align}
and denote by $\widehat{I}_A$ the generators in a representation on the fibres 
$E_x \cong \C^p$. By the invariant splitting, one has the following commutation 
relations:
\begin{align}
 \left[\widehat{I}_\alpha ,\widehat{I}_\beta \right]  = f_{\alpha \beta}^{\ \ 
\gamma} \widehat{I}_\gamma \; , \quad 
 \left[\widehat{I}_\alpha ,\widehat{I}_\mu \right]  = f_{\alpha \mu}^{\ \ 
\nu} \widehat{I}_\nu \; , \quad
 \left[\widehat{I}_\mu ,\widehat{I}_\nu \right]  = f_{\mu \nu}^{\ \ 
\alpha} \widehat{I}_\alpha + f_{\mu \nu}^{\ \ \sigma} \widehat{I}_\sigma \; , 
\label{eqn:Lie_algebra_su}
\end{align}
for $\alpha, \beta, \gamma =  2n+2, \ldots , (n+1)^2$ and $\mu,\nu,\sigma=  1, 
\ldots , 2n+1$. A suitable choice of these structure constants can be found 
in~\cite{Harland:2011zs,Ivanova:2012vz,Bunk:2014kva,Bunk:2014coa}.

 Next, we simplify the ansatz by demanding $X_{\hat{\mu}} = X_{\hat{\mu}}(t)$; 
i.e. not all connections $\Acal$ on $E$ can be reached anymore. Moreover, this 
demand is only valid in any trivialisation if the following 
conditions hold (see~\cite{Bunk:2014coa} for further details)
\begin{equation}
 \left[ \widehat{I}_\alpha , X_\mu \right] = f_{\alpha \mu}^{\ \ \nu} X_\nu 
\and \left[ \widehat{I}_\alpha , X_{2n+2} \right] = 0 \for \mu,\nu =1,\ldots, 
2n+1 \; . \label{eqn:equivariance}
\end{equation}
The $(f_{\alpha \mu}^{\ \ \nu})$ can be interpreted as the 
matrix elements $(\rho_*(I_\alpha))_\mu^{\ \nu}$ of a (suitably chosen) 
representation $\rho$ of $\surm(n)$ on the typical fibre of $T 
\mfd{2n+1}$. The representation theoretic content of~\eqref{eqn:equivariance} 
is that the matrix-valued functions $X_{\hat{\mu}}$ have to transform in a 
representation of $\surmL(n)$.

 Computing the curvature $\Fcal_\Acal$ for the 
ansatz~\eqref{eqn:ansatz_generic} together with the \emph{equivariance 
condition}~\eqref{eqn:equivariance} then yields
\begin{align}
 \Fcal_\Acal = &\Fcal_{\widehat{\Gamma}^P} + 
 \frac{1}{2} \left( \left[ X_a, X_b \right] + T_{a  \, b}^{2n+1} X_{2n+1} 
\right) e^{a} \wedge e^b 
+ \left(  \left[ X_a, X_{2n+1} \right] + T_{a \, 2n+1}^{b} X_{b} \right) e^a 
\wedge e^{2n+1} \notag \\*
&+\left( \left[X_a ,X_{2n+2}\right]  - \tfrac{\diff }{\diff t} X_a \right) e^a 
\wedge e^{2n+2} 
+ \left( \left[X_{2n+1} ,X_{2n+2}\right]  - \tfrac{\diff }{\diff t} X_{2n+1} 
\right) e^{2n+1} \wedge e^{2n+2} \; ,
\end{align}
with $\Fcal_{\widehat{\Gamma}^P} $ is the curvature of $\widehat{\Gamma}^P$, 
and $a,b=1,\ldots, 2n$.
 The HYM instanton equations~\eqref{eqn:holo_cond} and~\eqref{eqn:stab_cond} 
reduce for the ansatz to a set of matrix 
equations for the $X_{\hat{\mu}}$, which are given in~\cite{Ivanova:2012vz} 
(note that $X_{2n+2}=0$ for this case). Moreover, $\Fcal_{\widehat{\Gamma}^P} $ 
already satisfies the HYM equations, as the $\widehat{\Gamma}^P$ is the lift of 
an $\surm(n)$-instanton and the corresponding $\surm(n)$-principal bundle is 
a subbundle in the $\surm(n+1)$-principal bundle associated to $E$.

\paragraph{Matrix equations: real basis}
For completeness, the resulting matrix HINP-equations in the real basis 
$\{e^{\hat{\mu}} \}$ are the holomorphicity conditions
\begin{subequations}
\label{eqn:matrix_equations_real}
\begin{align}
  \left[ X_{2j-1} , X_{2k-1} \right] - \left[ X_{2j} , X_{2k} \right]&=0 \; , 
\\
\left[ X_{2j-1} , X_{2k} \right] + \left[ X_{2j} , X_{2k-1} \right] &= 0\; , 
\\
 \left[ X_{2j-1} , X_{2n+2} \right] + \left[ X_{2j} , X_{2n+1} \right] &=
\tfrac{\diff }{ \diff t} X_{2j-1} + \tfrac{n+1}{n} X_{2j-1}  \; ,\\
 \left[ X_{2j} , X_{2n+2} \right] - \left[ X_{2j-1} , X_{2n+1} \right] &=
\tfrac{\diff }{ \diff t} X_{2j} + \tfrac{n+1}{n} X_{2j}  \; ,
\end{align}
for $j,k = 1, \ldots , n$ and the stability-like condition
\begin{equation}
 \tfrac{\diff }{\diff t} X_{2n+1} + 2n X_{2n+1} = \sum_{k=1}^{n+1} \left[ 
X_{2k-1} , X_{2k} \right] \; .
\end{equation}
\end{subequations}
\paragraph{Matrix equations: complex basis}
For the intents and purposes here, it is more convenient to switch to the 
complex basis $\{\theta^j, \bar{\theta}^j\}$ defined 
in~\eqref{eqn:def_cplx_forms} and introduce
\begin{equation}
 Y_j \coloneqq \frac{1}{2} \left( X_{2j} - \im X_{2j-1} \right) \and 
 Y_{\bar{j}} \coloneqq \frac{1}{2} \left( X_{2j} + \im X_{2j-1} \right) \for 
j=1,2,\ldots ,n+1 \; . \label{eqn:def_Y-matrices}
\end{equation}
Hence, $Y_{\bar{j}} = - (Y_{j})^\dagger$ since $X_{\hat{\mu}}(t) \in 
\surmL(n+1)$ for all $t\in \R$.
For the $Y_j: \R \to \End(\C^p) $ one finds the holomorphicity 
conditions
\begin{subequations}
\label{eqn:matrix_equations}
\begin{equation}
  \tfrac{\diff }{\diff t} Y_j + \tfrac{n+1}{n} Y_j = 2  \left[Y_j , 
Y_{n+1}\right] \and  \left[Y_j , Y_{k}\right] =0  \for j,k = 1,\ldots,n \; ,
\end{equation}
and the adjoint equations thereof. The stability-like condition reads
\begin{equation}
 \tfrac{\diff }{\diff t} \left( Y_{n+1}^{\phantom{\dagger}} + 
Y_{n+1}^\dagger \right) + 2n \left( Y_{n+1}^{\phantom{\dagger}} + 
Y_{n+1}^\dagger \right) +2 \sum_{j=1}^{n+1} \left[Y_j^{\phantom{\dagger}} , 
Y_{j}^\dagger \right] =0\;.
\end{equation}
\end{subequations}
The equivariance conditions for the complex matrices are
\begin{equation}
 \left[\widehat{I}_\alpha , Y_j \right] =- \im f_{\alpha 2j-1}^{ \ \ 2j} Y_j 
\and \left[\widehat{I}_\alpha , Y_{n+1} \right] =0 \; ,  
\label{eqn:equivariance_cplx}
\end{equation}
for $ j,=1,\ldots,n$. For these calculations we have used the choice of 
structure constants $f_{\alpha \mu}^{\ \ \nu} = 0$ if $\mu$ or $\nu = 2n+1$ and 
$f_{\alpha a}^{\ \ b} \propto \omega_{ab}$, see for 
instance~\cite{Ivanova:2012vz,Bunk:2014kva,Bunk:2014coa}.
\paragraph{Change of trivialisation}
The remaining nontrivial effects of a change of trivialisation of the bundle 
$E$ over $\mathrm{Cyl}(\mfd{2n+1})$ are given by the set of functions $\{g(t) 
|\,g : \R \to \surm(p)\}$ that act as 
\begin{equation}
 X_\mu \mapsto \Ad(g) X_\mu \for \mu =1,\ldots,2n+1 \and X_{2n+2} \mapsto 
\Ad(g) X_{2n+2}  -\left( \tfrac{\diff}{\diff t} g \right) g^{-1} \; 
,\label{eqn:change_trivialisation}
\end{equation}
which follows from $\Acal \mapsto \Acal^g = \Ad(g) \Acal - (\diff g) g^{-1}$ 
and $g=g(t)$.~\footnote{We have simply replaced $g$ 
in~\eqref{eqn:trafo_connection} by $g^{-1}$.} Due to their adjoint 
transformation behaviour, the $X_\mu$ are 
sometimes called \emph{Higgs fields}, for example in quiver gauge theories. The 
inhomogeneous transformation of $X_{2n+2}$ is crucial to be able to ``gauge 
away'' this connection component. Furthermore, these gauge transformations (and 
their complexification) will be used to study the solutions of the matrix 
equations.
\paragraph{Yang-Mills with torsion}
The instanton equations (on the cone \emph{and} the cylinder) are equivalently 
given by 
\begin{equation}
  \star \Fcal_\Acal = - \frac{\omega^{n-1}}{(n-1)!} \wedge \Fcal_\Acal \; ,
  \label{eqn:instanton_alternative}
\end{equation}
where $\omega$ is the corresponding $(1,1)$-form ($\diff \omega =0$ on the 
cone, but $\diff \omega \neq 0$ on the cylinder). An immediate consequence is 
that the instanton equation for the integrable $\surm(n+1)$-structure implies 
the Yang-Mills equations, while this is not true for the $\surm(n+1)$-structure 
with torsion. In detail
\begin{subequations}
\begin{alignat}{5}
 &\text{cone: } & &\eqref{eqn:instanton_alternative}  & \quad &\Rightarrow  
\quad & &\diff_\Acal \star \Fcal_\Acal=0 & &\text{Yang-Mills} \; , \\
 &\text{cylinder: } & &\eqref{eqn:instanton_alternative}  & &\Rightarrow  &
&\diff_\Acal \star \Fcal_\Acal +\frac{\omega^{n-2}}{(n-2)!}\wedge \diff \omega 
\wedge \Fcal_\Acal =0 & \quad &\text{Yang-Mills with torsion} \; . 
\label{eqn:YM+torsion}
\end{alignat}
\end{subequations}
These \emph{torsionful Yang-Mills} equations~\eqref{eqn:YM+torsion}, which arise 
in the context of non-integrable $\G$-structures (with intrinsic torsion), have 
been studied in the literature 
before~\cite{Harland:2009yu,Harland:2010ix,Bauer:2010fia,Popov:2010rf,
Lechtenfeld:2012yw,Wolf:2012gz}. In particular, the torsion term does not 
automatically vanish on instantons because $\diff \omega$ contains $(2,1)$ 
and $(1,2)$-forms. 
This is, for instance, in contrast to the nearly Kähler case discussed 
in~\cite{Xu:2008}, in which nearly Kähler instantons were found to satisfy the 
ordinary Yang-Mills equations.

It is known that the appropriate functional for the torsionful Yang-Mills 
equations comprises the ordinary Yang-Mills functional plus an 
additional Chern-Simons term
\begin{equation}
 S_{\mathrm{YM+T}}(\Acal) = \int_{\mathrm{Cyl}(\mfd{2n+1})} \tr\left( 
\Fcal_\Acal 
\wedge \star \Fcal_\Acal \right) +\frac{\omega^{n-1}}{(n-1)!} \wedge 
\tr \left( \Fcal_\Acal \wedge \Fcal_\Acal \right) \; , \label{eqn:action}
\end{equation}
which is a gauge-invariant functional. The properties of $ S_{\mathrm{YM+T}}$ 
are the following: firstly and most importantly, instanton connections 
satisfying~\eqref{eqn:instanton_alternative} have $ S_{\mathrm{YM+T}}(\Acal) 
=0$, i.e. the \emph{action is finite}. Secondly, the stationary points 
of~\eqref{eqn:action} are the vanishing locus of the torsionful Yang-Mills 
equations (up to boundary terms). For this, we use $\Fcal_{\Acal + z \Psi} = 
\Fcal_\Acal + z \diff_\Acal \Psi + \frac{1}{2} z^2 \Psi \wedge \Psi$ for any 
$\Psi \in T_\Acal \mathbb{A}(E)$ and compute the variation
\begin{align}
 \delta S_{\mathrm{YM+T}}(\Acal) &\coloneqq \frac{\diff}{\diff z} 
S_{\mathrm{YM}}(\Acal + z \Psi ) \Big|_{z=0}  = \int_{\mathrm{Cyl}(\mfd{2n+1})} 
2 \ \tr \left(
\diff_\Acal \Psi 
\wedge \star \Fcal_\Acal \right) + 2 \frac{\omega^{n-1}}{(n-1)!} \wedge 
\tr \left( \Fcal_\Acal \wedge \diff_\Acal \Psi \right) \notag \\
&=2 \int_{\mathrm{Cyl}(\mfd{2n+1})} \tr \left(  \Psi \wedge \left( \diff_\Acal 
\star \Fcal_\Acal + \frac{\omega^{n-2}}{(n-2)!} \wedge 
\diff \omega \wedge  \Fcal_\Acal  \right) \right) \label{eqn:variation_action}\\
&\phantom{=2 \int_{\mathrm{Cyl}(\mfd{2n+1})} } +2\int_{\mathrm{Cyl}(\mfd{2n+1})} 
\diff \ \tr \left( \Psi \wedge \left( 
\star \Fcal_\Acal +  \frac{\omega^{n-1}}{(n-1)!}  
\wedge \Fcal_\Acal \right)  \right) \notag \; .
\end{align} 
The boundary term would vanish for closed manifolds. In our case, 
if one assumes 
$\mfd{2n+1}$ to be closed, the vanishing of the boundary term requires 
certain assumptions on the fall-off rate of $\Fcal_\Acal$ for $t\to \pm 
\infty$. Moreover, it is interesting to observe that the boundary term 
in~\eqref{eqn:variation_action} 
vanishes for instanton configurations.
\subsection{Rewriting the instanton equations}
\label{subsec:rewriting}
\paragraph{Real equations}
Returning to the instanton equations for the 
$X$-matrices~\eqref{eqn:matrix_equations_real}, the linear terms can be 
eliminated via a change of coordinates:
\begin{subequations}
\label{eqn:rescale_X}
\begin{alignat}{2}
 X_{2j-1} &\eqqcolon e^{-\tfrac{n+1}{n}t} \Xcal_{2j-1} \; , \qquad &  
 X_{2j}  &\eqqcolon e^{-\tfrac{n+1}{n}t} \Xcal_{2j}  \for j=1,\ldots,n  \; , \\
X_{2n+1}  &\eqqcolon e^{-2nt} \Xcal_{2n+1}  \;, \qquad &
X_{2n+2}  &\eqqcolon e^{-2nt} \Xcal_{2n+2} \; , 
\\
s &= -\frac{1}{2n} e^{-2nt} \in \R^-  \;, \qquad & \lambda_n(s) &\coloneqq 
\left( \frac{-1}{2ns} \right)^{2-\tfrac{n+1}{n^2}} \; .
\end{alignat}
\end{subequations}
Note that the exponent $2-\tfrac{n+1}{n^2}$ vanishes for $n=1$ and is strictly 
positive for any $n>1$. The matrix 
HINP equations~\eqref{eqn:matrix_equations_real} read now as follows:
\begin{subequations}
\label{eqn:matrix_equations_real_reformulated}
\begin{alignat}{3}
  \left[ \Xcal_{2j-1} , \Xcal_{2k-1} \right] - \left[ \Xcal_{2j} , 
\Xcal_{2k} \right] &= 0 &  &\and &
 \left[ \Xcal_{2j-1} , \Xcal_{2k} \right] + \left[ \Xcal_{2j} , 
\Xcal_{2k-1} \right] &=0 \; , \\
 \left[ \Xcal_{2j-1} , \Xcal_{2n+2} \right] + \left[ \Xcal_{2j} , 
\Xcal_{2n+1} \right]&= \tfrac{\diff }{ \diff s} \Xcal_{2j-1}   & &\and &
 \left[ \Xcal_{2j} , \Xcal_{2n+2} \right] - \left[ \Xcal_{2j-1} , 
\Xcal_{2n+1} \right] &= \tfrac{\diff }{ \diff s} \Xcal_{2j}    \; ,
\end{alignat}
for $j,k = 1,\ldots, n$ and 
\begin{equation}
 \tfrac{\diff }{\diff s} \Xcal_{2n+1}  = \lambda_n(s) \sum_{k=1}^{n} \left[ 
\Xcal_{2k-1} , \Xcal_{2k} \right]  + \left[ \Xcal_{2n+1} , \Xcal_{2n+2} 
\right]\; .
\end{equation}
\end{subequations}
\paragraph{Complex equations}
Completely analogous,  the change of coordinates for the complex equations is 
performed via
\begin{align}
 Y_j &\eqqcolon e^{-\tfrac{n+1}{n}t} \Ycal_j \for j=1,\ldots, n  \and 
Y_{n+1} \eqqcolon e^{-2nt} \Zcal \; . \label{eqn:rescaling}
\end{align}
We will refer to this set of matrices simply by $(\Ycal,\Zcal)$. In summary, 
the instanton equations are now comprised by the ``complex equations'' 
\begin{subequations}
\label{eqn:matrix_equations_reformulated}
\begin{align}
  \left[\Ycal_j , \Ycal_{k} \right] =0 
 \and \label{eqn:inst_Nahm} 
  \tfrac{\diff }{\diff s} \Ycal_j  = 2  \left[\Ycal_j , \Zcal \right] 
  \for j,k = 1,\ldots,n  \; ,
\end{align}
and the ``real equation''
\begin{equation}
 \tfrac{\diff }{\diff s} \left( \Zcal + \Zcal^\dagger \right) + 2 
\left[ \Zcal , \Zcal^\dagger \right]  +2 \lambda_n(s) \sum_{j=1}^{n} 
\left[\Ycal_j^{\phantom{\dagger}} , \Ycal_{j}^\dagger \right]=0  \;. 
\label{eqn:inst_stability}
\end{equation}
\end{subequations}
These equations are reminiscent to the considerations of the instantons on 
$\R^4\backslash \{0\}$ 
of~\cite{Donaldson:1984,Kronheimer:1989,Kronheimer:1990}, 
and, in fact, they reduce to the same system for $n=1$, but in general one a 
Calabi-Yau $2$-fold $\C^2 \slash \Gamma$. To see this, we 
recall~\cite{Boyer:2008} that all 
$3$-dimensional Sasaki-Einstein spaces are given by $S^3\slash \Gamma$, where  
$\Gamma$ is a finite subgroup of $\su$ (and commutes with $\uo\subset \su$) 
which acts freely and isometrically from the left on $S^3 \cong \su$.
\paragraph{Remarks}
The equivariance conditions for the rescaled matrices 
$\{\Xcal_{\hat{\mu}} \}$ or $(\{\Ycal_{j}\},\Zcal) $ are exactly the same 
as~\eqref{eqn:equivariance} or~\eqref{eqn:equivariance_cplx}, respectively.

Moreover, the rescaling has another salient feature: the matrices 
$\{\Xcal_{\hat{\mu}} \}$ or $(\{\Ycal_{j}\},\Zcal) $ (as well as 
their derivatives) are bounded (see for instance~\cite{Kronheimer:1989}); in 
contrast, the original connection components will develop a pole at the 
origin $r=0$. This will become apparent once the boundary conditions are 
specified. For further details, see Appendix~\ref{subsec:boundedness_matrices}.

In addition, we observe that the exponents on the 
rescaling~\eqref{eqn:rescale_X} reflect the torsion 
components~\eqref{eqn:torsion}. The choice of a flat ``starting 
point'' $\Gamma =0$ would lead to Nahm-type equations straight away, but 
solutions to the resulting matrix equations would not interpolate between any 
(non-trivial) lifted instanton from $\mfd{2n+1}$ and instantons on the 
Calabi-Yau space $C(\mfd{2n+1})$, 
cf.~\cite{Harland:2010ix,Ivanova:2012vz}.
\paragraph{Real gauge group}
The full set of instanton equations~\eqref{eqn:matrix_equations_reformulated} 
is invariant under the action of the gauge group
\begin{equation}
 \widehat{\Gcal} \coloneqq \left\{ g(s) | g: \R^- \to \urm(p) \right\} 
\; ,
\label{eqn:gauge_group_real}
\end{equation}
wherein the action is defined via
\begin{subequations}
 \label{eqn:instanton_gauge_transf}
\begin{align}
 \Ycal_{j} &\mapsto \Ycal_{j}^g \coloneqq \Ad(g ) \Ycal_{j} 
\for j =1, \ldots, n \; ,\\
\Zcal &\mapsto \Zcal^g \coloneqq \Ad(g) \Zcal -\frac{1}{2} 
\left( \frac{\diff}{\diff s} g \right) g^{-1} \; .
\end{align}
\end{subequations}
Note that only the real equation~\eqref{eqn:inst_stability} requires $g^{-1} = 
g^\dagger$ for it to be gauge invariant.
Moreover, one can always find a gauge transformation 
$g\in \widehat{\Gcal}$ such that $\Zcal^g= (\Zcal^g)^\dagger$ (Hermitian) or, 
equivalently, $X_{2n+2}^g=0$.

In summary, these properties follow from~\eqref{eqn:change_trivialisation} as 
the $X$-matrices are extensions to a connection. However, the gauge 
group~\eqref{eqn:gauge_group_real} still contains a nontrivial centre 
$ \{ g(s) | g(s) =\phi(s) \mathds{1}_{p\times p} \text{ with }
\phi: \R^- \to \uo \} $, such 
that~\eqref{eqn:change_trivialisation} corresponds to the quotient of 
$\widehat{\Gcal}$ 
by its centre.
\paragraph{Complex gauge group}
Moreover, the complex equations~\eqref{eqn:inst_Nahm} allow for an action of 
the complexified gauge group
\begin{equation}
 \widehat{\Gcal}^\C \equiv \left\{ g(s) \big| g: \R^- \to \glrm(p,\C) \right\} 
\; ,
\end{equation}
given by 
\begin{subequations}
\label{eqn:instanton_cplx_gauge_transf}
\begin{alignat}{2}
\Ycal_{k} &\mapsto \Ad(g) \Ycal_{k} \; , & 
\qquad 
\Ycal_{\bar{k}} &\mapsto \Ad((g^{-1})^\dagger) \Ycal_{\bar{k}} \; ,
\for k=1,\ldots,n \; ,\\
\Zcal &\mapsto \Ad(g) \Zcal - \tfrac{1}{2} \left( 
\tfrac{\diff}{\diff s} g  \right) g^{-1} \; ,  
& \qquad 
\bar{\Zcal} &\mapsto \Ad((g^{-1})^\dagger) \bar{\Zcal} + 
\tfrac{1}{2} (g^{-1})^\dagger \left( \tfrac{\diff}{\diff s} g^\dagger \right) 
\; .
\end{alignat}
\end{subequations}
The extension to $\widehat{\Gcal}^\C$-invariance for the holomorphicity 
conditions exemplifies the generic 
situation discussed in Section~\ref{subsec:HYM_intro}.
\paragraph{Equivariance condition}
Actually, one needs to be a bit more careful in considering these equations 
and their symmetries. Recall that we restrict ourselves to the matrices 
$X_{\hat{\mu}}$ which satisfy the equivariance 
conditions~\eqref{eqn:equivariance}. However, if the equivariance conditions 
are not invariant under the gauge 
transformations~\eqref{eqn:instanton_gauge_transf}, then a solution obtained 
by gauge transformation might not be equivariant anymore.

The real gauge transformations can be interpreted as change of basis on the 
fibres $E_x \cong \C^p$ or, more appropriately, change of trivialisation. Since 
the $\widehat{I}_\alpha$ are representations of 
the generators $I_\alpha$ on these fibres, the same transformation acts on them 
as well. In order to preserve the Lie algebra~\eqref{eqn:Lie_algebra_su}, all 
generators have to transform as
\begin{equation}
 \widehat{I}_A \mapsto \Ad(g)\widehat{I}_A \for g \in \widehat{\Gcal} \and A = 
1,\ldots 
(n+1)^2- 1 \; .
\end{equation}
The same transformation behaviour is adopted when passing to the 
complexified gauge group. This renders $[\widehat{I}_\alpha,X_\mu] = f_{\alpha 
\mu}^{\ \ \nu} X_\nu$ into a gauge invariant condition for both 
$\widehat{\Gcal}$ and 
$\widehat{\Gcal}^\C$-transformations; the $\widehat{\Gcal}^\C$-invariance 
follows 
as~\eqref{eqn:equivariance_cplx} does not intertwine $\{Y_k\}$ and 
$\{Y_{\bar{k}}\}$ for any $k=1,\ldots,n+1$. 
Unfortunately, $[\widehat{I}_\alpha,X_{2n+2}]=0$ is not gauge invariant, due 
the inhomogeneous transformation behaviour~\eqref{eqn:change_trivialisation} of 
$X_{2n+2}$. However, the way out is that we will only impose this 
last condition at the very end, i.e. once we have chosen a gauge transformation 
$g$ such that $X_{2n+2}^g=0$, the last equivariance condition follows trivially.
\paragraph{Boundary conditions}
We observe that a trivial solution 
of~\eqref{eqn:matrix_equations_real_reformulated} is 
\begin{equation}
 \Xcal_{2n+2}(s)=0 \and  \Xcal_\mu (s)= T_\mu \with [T_\mu,T_\nu]=0 \for 
\mu,\nu=1,\ldots,2n+1 \; ,
\end{equation}
where the (constant) $T_\mu$ are elements in the Cartan subalgebra of 
$\surmL(p)$; i.e. the (real) $(p{-}1)$-dimensional space spanned by the 
diagonal, traceless matrices with purely imaginary values. From the 
rescaling~\eqref{eqn:rescale_X} of the $X_{\hat{\mu}}$, it is apparent that 
these matrices become singular as $r\to 0$ ($t \to -\infty $ or $ s \to 
-\infty$).
Following~\cite{Kronheimer:1989,Hitchin:1991}, it is appropriate to choose the 
boundary conditions for 
$X_{\mu}$ to be\footnote{One does not need to worry about $X_{2n+2}$, 
as it can always be gauged away.}
\begin{subequations}
\label{eqn:aux_boundary_cond}
\begin{align}
 s &\to 0: \quad  X_\mu(s) \to 0 \for \mu=1,\ldots,2n+1 \and \\
  s &\to -\infty : \quad \exists g_0 \in \urm(p) \; \text{ such that } \; 
\Xcal_{\mu}(s) \to \Ad(g_0) T_{\mu}  \for \mu=1,\ldots,2n+1 \; .
\end{align}
\end{subequations}
One can show~\cite{Kronheimer:1989} that this implies the existence of the 
limit 
of $\Xcal_{\mu}$ for $s\to 0$. Hence, the solutions extend to the 
interval $(-\infty,0]$, see also Appendix~\ref{subsec:boundedness_matrices}. 
Thus, we are led to consider~\eqref{eqn:matrix_equations_real_reformulated} 
for matrices $\Xcal_\mu(s)$ over $(-\infty,0]$ with one remaining 
boundary condition: 
\begin{equation}
\label{eqn:boundary_cond}
 \exists g_0 \in \urm(p) \text{  such that } \forall 
\mu=1,\ldots,2n+1: \lim_{s\to 
-\infty}\Xcal_{\mu}(s) =\Ad(g_0) T_{\mu} \;.
\end{equation}
Moreover, since one has first order differential equations it suffices 
to impose this one boundary condition, here at $s=-\infty$. Thus, the values of 
$\Ycal_k$ at $s=0$ are completely determined by the solution. 
Following~\cite{Kronheimer:1989}, we
observe that~\eqref{eqn:inst_Nahm} implies that $\Ycal_{k}(s)$ lies entirely in 
a single adjoint orbit $\mathcal{O}(k)$ of the complex group 
$\widehat{\Gcal}^\C$, for 
each $k=1,\ldots,n$.
Next, assuming that $\T_{k} = \tfrac{1}{2} \left( T_{2k} - \im 
T_{2k-1} \right)$ for $k=1,\ldots,n$ is a \emph{regular tuple} in the Cartan 
subalgebra of $\widehat{\gfrak}^\C$ in the sense of~\cite{Kronheimer:1989}(that 
is the 
joint stabiliser of the $T_\mu$ in $\surm(p)$ is the maximal torus), one 
obtains that $\Ycal_{k}(s{=}0) \in \mathcal{O}(k)$, i.e. the 
values at $s=0$ are in a conjugacy class of $\T_{k}$. Moreover, only the 
conjugacy class has a gauge-invariant meaning.

Nonetheless, the boundary conditions~\eqref{eqn:boundary_cond} clearly show 
that the original connection~\eqref{eqn:ansatz_generic} develops the following 
poles at the origin $r=0$ of the Calabi-Yau cone:
\begin{equation}
 \lim_{r\to0} r^{\tfrac{n+1}{n}} X_a = \Ad(g_0)T_a \for a=1,\ldots,2n \and 
\lim_{r\to0} r^{2n} X_{2n+1} = \Ad(g_0)T_{2n+1} \; .
\end{equation}
Note that the case $n=1$ is reminiscent to the \emph{instantons with poles} 
considered in~\cite{Kronheimer:1989}.
% 
%%%%%%%%%%%%%%%%%%%%%%%%%%%%%%%%%%%%%%%%%%%%%%%%%%%%%%%%%%%%%%%%%%%%%%%%%%%%%%%
%%%%%%%%%%%%%%%%%%%%%%%%%%%%%%%%%%%%%%%%%%%%%%%%%%%%%%%%%%%%%%%%%%%%%%%%%%%%%%%
% 
\subsection{Geometric structure}
\label{subsec:Geo_structure}
\paragraph{Space of connections under consideration}
Consider the space of $\surmL(n+1)$-valued connections $\mathbb{A}(E)$ in 
which any element can be parameterised as in~\eqref{eqn:ansatz_generic}. Due to 
the ansatz of Section~\ref{subsec:Ansatz}, we restrict ourselves to the 
subspace $\mathbb{A}_{\mathrm{equi}}(E) \subset \mathbb{A}(E)$ of 
connections which satisfy~\eqref{eqn:equivariance}. Specialising the 
considerations of Section~\ref{subsec:HYM_intro}, we will now establish certain 
(formal) geometric structures.
\paragraph{Kähler structure}
The first step is to establish a Kähler structure on 
$\mathbb{A}_{\mathrm{equi}}(E)$. Since $ \mathbb{A}_{\mathrm{equi}}(E) $ 
descends from the space of all connection $ \mathbb{A}(E)$, one can simply 
obtain the geometric structures by restriction. A tangent vector
\begin{equation}
 \boldsymbol{y}= \sum_{j=1}^{n+1} \left( \boldsymbol{y}_j \theta^j + 
\boldsymbol{y}_{\bar{j}}  \bar{\theta}^j \right) 
\end{equation}
at a point $\Acal\in\mathbb{A}_{\mathrm{equi}}(E)$ is defined by the 
linearisation 
of~\eqref{eqn:matrix_equations} for paths $\boldsymbol{y}_{j}:\R \to 
\surmL(p)$. Their gauge transformations are 
\begin{equation}
 \boldsymbol{y}_{j} \to \boldsymbol{y}_{j}^{g} \coloneqq \Ad(g) 
\boldsymbol{y}_{j} \for j = 1,\ldots,n+1\;.
\end{equation}
Taking the generic expressions for the metric~\eqref{eqn:metric_connections} 
and the symplectic structure~\eqref{eqn:symplectic_form_connections}, we can 
specialise to the case at hand by transition to the cylinder and neglecting the 
volume integral of $\mfd{2n+1}$. Thus, for a metric on 
$\mathbb{A}_{\mathrm{equi}}$ we obtain
\begin{equation}
 \boldsymbol{g}_{|\Acal}(\boldsymbol{y}^{(1)}, \boldsymbol{y}^{(2)})\equiv 
2 \int_{\R} \diff t \ e^{2nt} \; \tr 
\left\{ \sum_{j=1}^{n+1} \left( \boldsymbol{y}_{j}^{(1)\dagger} 
\boldsymbol{y}_{j}^{(2)}  + 
\boldsymbol{y}_{j}^{(1)} \boldsymbol{y}_{j}^{(2)\dagger}  \right) 
\right\}\; .
\label{eqn:metric_matrices}
\end{equation}
Similarly, the symplectic form reads as
\begin{equation}
  \boldsymbol{\omega}_{|\Acal}(\boldsymbol{y}^{(1)}, \boldsymbol{y}^{(2)}) 
\equiv  
  -2\im \int_{\R} \diff t \ e^{2nt} \; \tr 
\left\{ \sum_{j=1}^{n+1} \left( \boldsymbol{y}_{j}^{(1)\dagger} 
\boldsymbol{y}_{j}^{(2)}  - 
\boldsymbol{y}_{j}^{(1)} \boldsymbol{y}_{j}^{(2)\dagger}  \right)  
\right\} \; .
\label{eqn:symplectic_matrices}
\end{equation}
Moreover, a complex structure $\boldsymbol{J}$ on 
$\mathbb{A}(E)_{\mathrm{equiv}}$ has been given 
in~\eqref{eqn:cplx_str_connections}. Keeping in mind 
that~\eqref{eqn:def_cplx_forms} implies $J = -J_{\mathrm{can}}$, we obtain
\begin{equation}
 \boldsymbol{J}_{|\Acal}(\boldsymbol{y}) = J(\boldsymbol{y}) = 
\im \sum_{j=1}^{n+1} \left( \boldsymbol{y}_j \theta^j - 
\boldsymbol{y}_{\bar{j}}  \bar{\theta}^j \right)
\end{equation}
As before, the symplectic form $\boldsymbol{\omega}$ and the metric 
$\boldsymbol{g}$ 
are compatible, i.e $\boldsymbol{g}(\boldsymbol{J} \cdot,  \cdot ) = 
\boldsymbol{\omega}(\cdot, \cdot)$. We note that both structures are 
gauge-invariant by construction.
\paragraph{Moment map}
The subspace of holomorphic connections $\mathbb{A}^{1,1}_{\mathrm{equi}}(E) 
\subset \mathbb{A}_{\mathrm{equi}}(E)$ is defined by the 
condition~\eqref{eqn:inst_Nahm}. 
This condition only restricts the allowed endmorphism-valued $1$-forms, because 
$\widehat{\Gamma}^P$ is already a $(1,1)$-type connection, 
since it is an HYM-instanton.
Again, the metric $\boldsymbol{g}$ and Kähler form $\boldsymbol{\omega}$ 
descend to $\mathbb{A}^{1,1}_{\mathrm{equi}}(E)$ from the corresponding objects 
on $\mathbb{A}_{\mathrm{equi}}(E)$. Moreover, on the Kähler space 
$\mathbb{A}^{1,1}_{\mathrm{equi}}(E)$, one defines a moment map
\begin{equation}
\begin{split}
 \mu :\qquad \mathbb{A}^{1,1}_{\mathrm{equi}}(E) &\to \widehat{\gfrak}_0= 
\mathrm{Lie}(\widehat{\Gcal}_0) \\
\left(\Ycal, \Zcal \right) &\mapsto  \im \left(
\frac{\diff}{\diff s} \left( \Zcal + \Zcal^\dagger \right) + 
2 \left[ \Zcal  , \Zcal^\dagger \right] + 2 \ \lambda_n(s)
\sum_{k=1}^n \left[ \Ycal_{k}^{\phantom{\dagger}} , \Ycal_{k}^\dagger \right]  
\right)
\; ,
\end{split} \label{eqn:moment_map}
\end{equation}
where $\widehat{\Gcal}_0$ is the corresponding \emph{framed} gauge group. That 
is
\begin{equation}
 \widehat{\Gcal}_0 \coloneqq \left\{ g(s) | g: \R^- \to \urm(p) \; ,  \; 
\lim_{s\to0}g(s)=\lim_{s\to -\infty}g(s)=1 \right\} \; . 
\label{eqn:gauge_group_framed}
\end{equation}
It is an important realisation that on the non-compact Calabi-Yau cone (and the 
conformally equivalent cylinder) one has to compensate the appearing boundary 
terms in Stokes' theorem by the transition to the framed gauge transformations. 
The details of the proof that \eqref{eqn:moment_map} satisfies 
conditions~\eqref{eqn:aux_moment_map} are given in the 
Appendix~\ref{subsec:proof_moment_map}. Here, we just note that the 
map~\eqref{eqn:moment_map} maps the matrices $\left(\Ycal, \Zcal \right)$ into 
the correct space: the factor of $\im$ renders the expression anti-hermitian; 
while the boundary conditions~\eqref{eqn:aux_boundary_cond} together with the 
gauge choice $\Zcal = -\Zcal^\dagger$ yield the vanishing of $\mu\left(\Ycal, 
\Zcal \right)$  at $s\to0$ and $s\to-\infty$. 

The part of instanton moduli space that is connected with the lift 
$\widehat{\Gamma}^P$ (in the sense of our 
ansatz~\eqref{eqn:ansatz_generic}) is then readily obtained by the Kähler 
quotient 
\begin{equation}
 \mathcal{M}_{\Gamma^P} = \mu^{-1} (0) \slash \widehat{\Gcal}_0 \; .
\end{equation}
\paragraph{Stable points}
Alternatively, one can describe this part of the moduli space via the 
stable points
\begin{equation}
\mathbb{A}^{1,1}_{st}(E) \equiv \left\{ \widehat{\Gamma}^P + X
\in \mathbb{A}^{1,1}(E)  \big| \; (\widehat{\Gcal}_0^\C)_{( \Ycal, \Zcal )} 
\cap 
\mu^{-1}(0)\neq \emptyset  \right\} \; ,
\end{equation}
where the tuple $(\Ycal,\Zcal)$ is obtained from $X$ via complex linear 
combinations and rescaling as before. The moduli space arises then by taking 
the $\widehat{\Gcal}_0^\C$-quotient
\begin{equation}
 \mathbb{A}^{1,1}_{st}(E) \slash \widehat{\Gcal}_0^\C \cong 
\mathcal{M}_{\Gamma^P} \; .
\end{equation}
We argue in the next couple of paragraphs that it suffices 
to solve the complex equations~\eqref{eqn:inst_Nahm}, because the 
solution to the real equation~\eqref{eqn:inst_stability} follows from a framed 
complex 
gauge transformation. More precisely: for 
every point in $\mathbb{A}^{1,1}_{\mathrm{equi}}(E)$ there exists a unique 
point in the complex gauge orbit such that the real equation is satisfied. In 
other words, every point in $\mathbb{A}^{1,1}_{\mathrm{equi}}(E)$ is stable.
\subsection{Solutions to matrix equations}
\label{subsec:solution}
\paragraph{Solutions to complex equation}
In the spirit of~\cite{Donaldson:1984}, one can also understand the  
complex equations as being locally trivial.  That is, 
take~\eqref{eqn:instanton_cplx_gauge_transf} and demand the gauge transformed 
$\Zcal$ to be zero
\begin{equation}
 \Zcal^g = \Ad(g) \Zcal - \tfrac{1}{2} \left( \tfrac{\diff}{\diff 
s} g\right)  g^{-1}   \stackrel{!}{=} 0 \qquad  \Rightarrow  \qquad 
\Zcal = \tfrac{1}{2} g^{-1} \tfrac{\diff}{\diff s} g \; .
\end{equation}
From the holomorphicity equations~\eqref{eqn:inst_Nahm} one obtains
\begin{equation}
 \tfrac{\diff}{\diff s} \Ycal_{k}^g =0 \and \Ycal_{k}^g = 
\Ad(g_0) \mathcal{T}_{k} \;  \text{ with } \quad 
\left[\T_{j},\T_{k} \right]=0\; ,
\end{equation}
for $j,k=1,\ldots,n$ and $g_0$ is a \emph{constant} gauge 
transformation\footnote{This $g_0$ can also be gauge away to $1$.}. 
Consequently, the general local solution of the complex 
equations~\eqref{eqn:inst_Nahm} is
\begin{equation}
 \quad \Ycal_{k} = \Ad(g^{-1}) \T_{k} \;  \text{ with 
} \quad \left[\T_{j},\T_{k} \right]=0 \and 
\Zcal=\frac{1}{2} g^{-1} \frac{\diff}{\diff s} g  \; , 
\label{eqn:solution_complex}
\end{equation}
for \emph{any} $g \in \widehat{\Gcal}^\C$.
A solution to the commutator constraint is choosing $\T_{k}$ for 
$k=1,\ldots,n$ as elements of the Cartan subalgebra of the Lie algebra 
$\glrmL(p,\C)$, which are all diagonal (complex) $p\times p$ matrices.
\paragraph{Solution to the real equation}
In any case, one can in principle solve the complex equations; now, the real 
equation~\eqref{eqn:inst_stability} needs to be solved as well. Following 
the ideas of~\cite{Donaldson:1984}, the considerations 
are split in two steps: (i) a variational description and (ii) a differential 
inequality. We provide the details of (i) in this paragraph, while we 
postpone the details of (ii) to the Appendix~\ref{sec:proof}.
 Let us recall that the complete 
set of instanton equations is gauge-invariant under $\widehat{\Gcal}$. Thus, 
define for 
each $g \in \widehat{\Gcal}^\C$ the map
\begin{equation}
 h = h(g) =  g^\dagger g : \R^- \to \glrm(p,\C) \slash \urm(p)\;.
\end{equation}
The quotient $ \glrm(p,\C) \slash \urm(p)$ can be identified with the set of 
positive, self-adjoint $p\times p$ matrices.
Then, fix a tuple $(\Ycal,\Zcal)$ and define the functional 
$\mathcal{L}_{\epsilon}[g]$ for $g$
\begin{equation}
 \mathcal{L}_{\epsilon}[g] = \tfrac{1}{2} 
\int_{-\tfrac{1}{\epsilon}}^{-\epsilon} \diff s \ \tr \left( \left| 
\Zcal^g +  (\Zcal^\dagger)^g \right|^2 + 2 \lambda_n(s) \sum_{k=1}^{n} \left| 
\Ycal_{k}^g \right|^2 \right) \for 0<\epsilon<1 \; ,
\label{eqn:Lagrange_functional}
\end{equation}
 where $(\Ycal^g,\Zcal^g)$ denotes the gauge-transformed tuple.
For the variation of~\eqref{eqn:Lagrange_functional} it suffices to 
consider variations with $\delta g = \delta g^\dagger$ around $g=1$, but of 
course $\delta g \neq 0$.
Then the gauge transformations~\eqref{eqn:instanton_cplx_gauge_transf} imply
\begin{equation}
 \delta \Zcal = \left[ \delta g , \Zcal \right] - \tfrac{1}{2} 
\frac{\diff}{\diff s} \delta g \and \delta 
\Ycal_{k} = \left[ \delta g , \Ycal_{k} \right] \for k=1,\ldots,n \; .
\end{equation}
The variation then leads to
\begin{equation}
 \delta_g \mathcal{L}_{\epsilon} =  -\im
\int_{-\tfrac{1}{\epsilon}}^{-\epsilon} 
 \diff s \ \tr \left\{ \mu(\Ycal,\Zcal) \ \delta g 
\right\} \; ,
\end{equation}
i.e. critical points of~\eqref{eqn:Lagrange_functional} are precisely the
zero-level set of the moment map.
Next, we take the solution~\eqref{eqn:solution_complex} and 
insert it as a starting point for $\mathcal{L}_{\epsilon}$. Thus, one obtains 
a functional for $h$
\begin{align}
 \mathcal{L}_{\epsilon} [h] &=  \tfrac{1}{2} 
\int_{-\tfrac{1}{\epsilon}}^{-\epsilon} \diff s \ \left\{ \tfrac{1}{4} \tr 
\left( h^{-1} \frac{\diff h}{\diff s} \right)^2  + 2 \lambda_n(s) \sum_{k=1}^n 
\tr\left( h \T_{k} h^{-1}  \T_{k}^\dagger \right)\right\} \\
&=  \tfrac{1}{2} 
\int_{-\tfrac{1}{\epsilon}}^{-\epsilon} \diff s \ \left\{ \tfrac{1}{4} \tr 
\left( h^{-1} 
\frac{\diff h}{\diff s} \right)^2  + V \right\} \notag \; .
\end{align}
Following~\cite{Donaldson:1984}, the potential $V(h)= 2 \lambda_n(s) 
\sum_{k=1}^n \tr\left( h \T_{k} h^{-1}  \T_{k}^\dagger \right)$ is 
positive\footnote{Note that $\lambda_n(s)$ is strictly positive and 
smooth on $\left(-\tfrac{1}{\epsilon},-\epsilon \right)$ for any $0<\epsilon 
<1$.}, implying that for \emph{any} boundary values $h_- , h_+ \in 
\glrm(p,\C) \slash \urm(p)$ there exists a continuous path\footnote{See for 
instance the note under~\cite[Corollary 2.13]{Donaldson:1984}: One knows that 
$\glrm(p,\C) \slash \urm (p)$ satisfies all necessary conditions for the 
existence of a unique stationary path between any two points.}
\begin{equation}
 h : \left[-\tfrac{1}{\epsilon},-\epsilon \right] \to  \glrm(p,\C) \slash 
\urm(p) \with h(-\tfrac{1}{\epsilon}) =h_- \and h(-\epsilon) = h_+ \; ,
\end{equation}
which is smooth in $I_\epsilon = \left(-\tfrac{1}{\epsilon},-\epsilon 
\right)$ and minimising the functional. Hence, for any choice of gauge 
transformation $g$ such that $g^\dagger g=h$ one has that $\left( \{ \T_{k} 
\}_{k=1,\ldots,n},0 \right)^g = \left( \{ \Ad(g)\T_{k}  
\}_{k=1,\ldots,n},-\tfrac{1}{2} (\tfrac{\diff}{\diff s}g ) g^{-1}  \right) $ 
satisfies the real equation in $I_\epsilon $ for any $0<\epsilon <1$. From 
now on, we restrict the attention to $h_+=h_-=1$, i.e. $h$ is framed.

The uniqueness of the solution $h$ on each interval $I_\epsilon$ 
and the existence of the limit $h_\infty$ for $\epsilon \to 0$ follows from the 
aforementioned differential inequality similar to~\cite{Donaldson:1984} and the 
discussion of~\cite[Lemma 3.17]{Kronheimer:1989}. The details are presented in 
Appendix~\ref{sec:proof}. The relevant (framed) gauge transformation is then 
simply given by $g = \sqrt{h_\infty}$.~\footnote{We use the \emph{unique} 
principal root of the positive Hermitian matrix $h$, which is a continuous 
operation. Consequently, the framing of $h$ implies the framing of 
$g$.}

However, we need to emphasise two crucial points. Firstly, the construction of  
a solution for the limit $\epsilon \to 0$ relies manifestly on the use of the 
boundary conditions~\eqref{eqn:boundary_cond}, and the fact that these give 
rise to a (constant) solution of \emph{both} the complex equations \emph{and} 
the real equation. Secondly, the corresponding complex gauge transformation 
$g=g(h_\infty)$ is only determined up to unitary gauge transformations, i.e. it 
is not unique. This ambiguity in the choice of $g$ can be removed, when we
recall that a $\widehat{\Gcal}$ gauge transformation suffices to eliminate 
$X_{2n+2}$. Hence, one can demand that the gauge-transformed system 
$(\Ycal^{g},\Zcal^{g})$ of a solution $(\Ycal,\Zcal)$ satisfies $\Zcal^{g} = 
(\Zcal^{g})^\dagger$. This fixes $g=g(h)$ uniquely, see also 
Appendix~\ref{subsec:limit} for further details.
\paragraph{Result}
In summary, it is sufficient to search for 
solutions $( \Ycal',\Zcal')$ of the complex equations~\eqref{eqn:inst_Nahm} on 
the interval $(-\infty,0]$ such that the boundary 
conditions~\eqref{eqn:boundary_cond} are satisfied. Then one has the existence 
of a unique complex gauge transformation $g$ such that 
\begin{enumerate}[(i)]
 \item $(\Ycal,\Zcal) = ( \Ycal',\Zcal')^g$ 
satisfies~\eqref{eqn:inst_stability} ,
\item $\Zcal$ is Hermitian (i.e. $\Xcal_{2n+2}=0$) and 
\item $g$ is bounded and framed.
\end{enumerate}
 In other words, it suffices to solve the complex equations 
subject to some boundary conditions and the real equation will be satisfied 
automatically.

Moreover, the above indicates that any point in 
$\mathbb{A}_{\mathrm{equi}}^{1,1}$ is stable, which we recall to be exactly the 
condition that every complex gauge orbit intersects $\mu^{-1}(0)$. We believe 
that this circumstance holds because we restricted ourselves to the space of 
equivariant connections. The benefit 
is then, that one, in principle, only has to show the solvability of the 
holomorphicity conditions in order to solve the instanton (matrix) equations. 
Nevertheless, one still has to find an ansatz that satisfies the equivariance 
conditions~\eqref{eqn:equivariance}.

%% file: remarks.tex
\subsection{Further directions}
Before concluding we can further exploit the results collected so far as well 
as illustrate another viewpoint of the HINP matrix equations.
\paragraph{Relation to coadjoint orbits}
Let us denote by $\Mcal_n(E)$  the moduli space of solutions to the 
complex and real equations satisfying the boundary 
conditions~\eqref{eqn:boundary_cond} together with the equivariance condition.
From the considerations above, we can establish the following map\footnote{I 
thank Richard Szabo for pointing this out to me.}
\begin{equation}
\label{eqn:map}
 \begin{split}
  \Mcal_n(E) &\to \mathcal{O}_{\T_1} \times \cdots \times 
\mathcal{O}_{\T_n} \\
(\Ycal,\Zcal) &\mapsto (\Ycal_1(0),\ldots,\Ycal_n(0))
 \end{split}
\end{equation}
where $\mathcal{O}_{\T_k} $ denotes the adjoint orbit of $\T_k$  in 
$\glrmL(p,\C)$.
Analogous to~\cite{Kronheimer:1989}, this map is a bijection due to the 
construction of the local solution~\eqref{eqn:solution_complex} and the 
uniqueness of the corresponding solution of the real and complex equations.   
Moreover, one knows that the orbit of an element $\T_k$ of the Cartan 
subalgebra is of the form $\glrm(p,\C)/ \mathrm{Stab}(\T_k) $. The 
product of coadjoint orbits in~\eqref{eqn:map} is a complex symplectic manifold 
of complex dimension $n \
\dim(\glrm(p,\C)) - \sum_{j=1}^n \dim(\mathrm{Stab}(\T_j))$. Each orbit is 
equipped with the so-called Kirillov-Kostant-Souriau symplectic form and the 
product thereof gives the symplectic structure on the total space. In 
addition, the bijection above preserves the holomorphic symplectic structure.
\paragraph{Relation to quiver representations}
The HINP matrix equations can be seen to define quiver representations, 
depending on the chosen $\surm(n{+}1)$-representation on the typical 
fibre $\C^p$. Then, by the employed ansatz, we decompose this representation 
with respect to $\surm(n)$ into 
\begin{equation}
 \C^p \Big|_{\surm(n)} = \bigoplus_{w \in J} \C^{n_w} \; ,
\end{equation}
where $\C^{n_w}$ carries a $n_w$-dimensional irreducible 
$\surm(n)$-representation. More explicitly, $w$ should be understood as pair of 
labels: let $\phi$ label the irreducible $\surm(n)$-representations and recall 
that the centraliser of $\surm(n)$ inside $\surm(n{+}1)$ is a $\uo$. Then each 
representation space $\C^{n_w}$ carries also a $\uo$-representation 
characterised by a ``charge'' $q$. Therefore, the decomposition is labelled by 
pairs $w=(\phi,q)$.

As a consequence, the equivariance 
condition~\eqref{eqn:equivariance} dictates the decomposition of the 
$X_\mu$-matrices into homomorphisms
\begin{equation}
X_\mu = \bigoplus_{w,w' \in J} (X_\mu)_{w,w'} \with (X_\mu)_{w,w'} \in 
\Hom\left(\C^{n_w},\C^{n_{w'}}\right) \; .
\end{equation}
The quiver representation then arises as follows: the set $Q_0$ of vertices is 
the set $\left\{ \C^{n_{w}} | w \in J \right\}$ of vector spaces  and the set 
$Q_1$ of arrows is given by the non-vanishing homomorphisms $\{ (X_\mu)_{w,w'} | 
w,w' \in J \; , \; \mu =1,\ldots, 2n+1\}$.

The instanton equations (or HINP equations) then lead to relations on the 
quiver representation. Examples for the arising quiver diagrams as well as 
their relations for the case $n=1$ and $\mfd{3}=S^3$ can be found 
in~\cite{Lechtenfeld:2014fza} and for $n=2$ and $\mfd{5}=S^5$
in~\cite{Lechtenfeld:2015ona}. To study the representations of a quiver one 
would rather use the constructions 
of~\cite{Lechtenfeld:2014fza,Lechtenfeld:2015ona}, instead of 
the ansatz employed here. Because once the bundle $E$ and the action of 
$\surm(n{+}1)$ on the fibres is chosen, there is no freedom to change the 
quiver representation anymore.

%% file: discussion.tex
\section{Conclusions}
\label{sec:conclusion}
It is known that the instanton moduli space over a Kähler manifold is a 
Kähler space. Therefore, also the moduli space of certain invariant connections 
should inherit this property. The overall situation remains unknown.

In the ansatz presented here, we restricted ourselves to a subset of all 
possible 
connections by, firstly, imposing an equivariance condition and 
simplifying to $t$-dependence only and by, secondly, fixing an instanton 
$\Gamma^P$ as a starting point. 
Hence, by this construction one can only reach a particular part of the full 
instanton moduli space by the solutions of the HINP matrix equations.

The arguments presented in this paper show that the reduced HINP matrix 
equations can be 
treated similarly to the Nahm-equations of $\su$ monopoles. As a consequence, 
one gains local solvability of the holomorphicity conditions together with the 
fact that any solution can be uniquely gauge-transformed into a solution of the 
stability-like condition. Moreover, the structure of the (framed) moduli space 
shares, at least locally, all features of a Kähler space due to the Kähler 
quotient construction or the GIT quotient. 

It is of interest to extend the ansatz presented here from cones to their 
smooth resolutions as in~\cite{Correia:2010,Ivanova:2012vz}, and to consider 
quiver gauge theories which can be associated to Calabi-Yau cones along the 
lines of~\cite{Lechtenfeld:2014fza,Lechtenfeld:2015ona}.
\section*{Acknowledgements}
I would like to thank Jakob Geipel, Lutz Habermann, Olaf Lechtenfeld and 
Alexander Popov for discussions and useful comments. This work was done within 
the framework of the DFG project LE 838/13. Moreover, I am grateful to the 
Research Training Group GRK 1463 ``Analysis, Geometry and String Theory'' for 
support.

%% file: appendix.tex
\section{Details}
In this appendix, we provide the proofs of the statements made in 
Sections~\ref{subsec:rewriting}--\ref{subsec:solution}. Although the steps 
are similar to those performed 
in~\cite{Donaldson:1984,Kronheimer:1989,Kronheimer:1990}, we believe that these 
are necessary because the HINP equations are generalisations of the Nahm 
equations.
\subsection{Boundedness of rescaled matrices}
\label{subsec:boundedness_matrices}
Recall the boundary conditions~\eqref{eqn:boundary_cond} for the original 
matrices
\begin{subequations}
\begin{alignat}{2}
 &t \to +\infty :& \qquad X_\mu &\to 0 \; ,\\
 &t \to -\infty :& \qquad e^{\tfrac{n+1}{n}t}X_a &\to \Ad(g_0) T_a \and 
e^{2nt}X_{2n+1} \to \Ad(g_0) T_{2n+1} \; .
\end{alignat}
\end{subequations}
Evaluating the asymptotic behaviour for $t \to + \infty$ 
of~\eqref{eqn:matrix_equations_real}, one finds the leading behaviour of (the 
real and imaginary part) of each matrix element to be
\begin{subequations}
\begin{alignat}{3}
 \frac{\diff}{\diff t} (X_a)_{AB} + \frac{n+1}{n}(X_a)_{AB} &\simeq 0 & 
\qquad &\rightarrow  \qquad & (X_a)_{AB} &\sim e^{-\tfrac{n+1}{n}t} \text{ as } 
t \to \infty \; ,\\
 \frac{\diff}{\diff t} (X_{2n+1})_{AB} + 2n(X_{2n+1})_{AB} &\simeq 0 &
\qquad &\rightarrow  \qquad & (X_{2n+1})_{AB} &\sim e^{-2nt} \text{ as } t \to 
\infty \; ,
\end{alignat}
\end{subequations}
because the commutator terms vanish faster than linear order.
These results imply the following:
\begin{enumerate}[(i)]
 \item The rescaled matrices $\Xcal_{\mu}$ of~\eqref{eqn:rescale_X} are bounded 
for $s\to 0$.
\item The commutators $e^{\tfrac{n+1}{n}t}[X_a,X_{2n+1}] $ are integrable 
over $(0,\infty) $.
\item The derivatives $\frac{\diff }{\diff t} \left( e^{\tfrac{n+1}{n}t} X_a 
\right)$ and $\frac{\diff }{\diff t} \left( e^{2nt} X_{2n+1} \right)$ are 
integrable, which follows by the use of the 
equations~\eqref{eqn:matrix_equations_real}.
\end{enumerate}
In conclusion, the $\Xcal_{\hat{\mu}}$ as well as their derivatives 
are bounded.
\subsection{Well-defined moment map}
\label{subsec:proof_moment_map}
We need to prove~\eqref{eqn:aux_moment_map} for $\mu$ defined 
in~\eqref{eqn:moment_map}; recall that $\mu(\Acal) := \Fcal_\Acal \wedge 
\frac{\widehat{\omega}^{n-1}}{(n-1)!}$ and we identified $\mu^*$ with $\mu$. 
Moreover, it is crucial to use the closed Kähler $2$-form from the cone, i.e. 
$\widehat{\omega}=e^{2t} \widetilde{\omega} $ on the cylinder. We will work with 
the original connection components $Y_k$ defined in~\eqref{eqn:def_Y-matrices}.

For the left-hand-side we proceed as follows:
Let $\phi \in \widehat{\mathfrak{g}}_0 $ and $\Psi= \Psi_k \theta^k - 
\Psi_k^\dagger \bar{\theta}^k $ be a tangent vector at $\Acal$. The 
duality pairing of Lie- and dual Lie-algebra is realised by the integration over 
the cylinder and the subsequent invariant product on $\mathfrak{u}(p)$. 
\begin{subequations}
\begin{align}
 (\phi,\Diff \mu_{|\Acal} ) \Psi &= 
\int_{\mathrm{Cyl}(\mfd{2n+1})} \tr \left\{ \phi 
\frac{\diff}{\diff z} \Fcal_{\Acal + z \Psi} \big|_{z=0}\right\} 
\wedge  \frac{\widehat{\omega}^n}{n!} \\
&= \int_{\mathbb{R}} \diff t \ e^{2nt} \ \tr  \Bigg\{ \phi \cdot \im 
\Bigg[ \frac{\diff}{\diff t} (\Psi_{n+1} + 
\Psi_{n+1}^\dagger) + 2n (\Psi_{n+1} + \Psi_{n+1}^\dagger)  \\*
&\qquad \qquad \qquad \qquad \qquad  + 2 \sum_{k=1}^{n+1}\left( \left[\Psi_k, 
Y_k^\dagger \right] + \left[Y_k, \Psi_k^\dagger \right] \right) \Bigg]
 \Bigg\} \cdot 
\int_{\mfd{2n+1}} \mathrm{vol} \; .\notag
\end{align}
\end{subequations}
Hence, for the dual moment map one can neglect the volume integral over 
$\mfd{2n+1}$ and the dual pairing is defined via the first integral over $t$. 

The compute the right-hand-side of~\eqref{eqn:aux_moment_map} we need to take a 
step back and derive the symplectic form on $\mathbb{A}$ 
from~\eqref{eqn:symplectic_form_connections} as follows
\begin{subequations}
\begin{align}
 \boldsymbol{\omega}_{|\Acal}(\Psi,\Xi) &= -\int_{\mathrm{Cyl}(\mfd{2n+1})} 
\tr\left(\Psi \wedge \Xi \right) \wedge \frac{\widehat{\omega}^n}{n!} \\
&= -2 \im \int_{\R} \diff t \ e^{2n t} \ \tr \sum_{k=1}^{n+1} 
\left\{ \Psi_k^\dagger \Xi_k - \Psi_k \Xi_k^\dagger \right\} \cdot 
\int_{\mfd{2n+1}} \mathrm{vol} \; .
\end{align}
\end{subequations}
Again, we can drop the volume of the Sasaki-Einstein space. Next, we need the 
infinitesimal gauge transformation generated by an (framed) Lie-algebra 
element $\phi$. From~\eqref{eqn:instanton_gauge_transf} we obtain
\begin{align}
 \phi^{\#}=\frac{\diff}{\diff z} Y_j^{g=\exp(z \phi)} \bigg|_{z=0} = 
\begin{cases} \left[ \phi , Y_j  \right] & ,j= 1,\ldots ,n     \\ \left[ \phi , 
Y_{n+1}\right]  - \frac{1}{2} \frac{\diff}{\diff t} \phi & , j=n+1 \; ,
\end{cases} 
\end{align}
which then brings us to
\begin{subequations}
\begin{align}
 \iota_{\phi^{\#}} \boldsymbol{\omega}_{|\Acal} (\Psi) &=  -2 \im 
\int_{\mathbb{R}} \diff t \ e^{2n t} \ \tr \Bigg\{ \sum_{k=1}^{n} 
\left\{ \left[ \phi , Y_k  \right]^\dagger \Psi_k - \left[ \phi , Y_k  
\right] \Psi_k^\dagger \right\}  \\
&\qquad \qquad  \qquad   +\left( \left[ \phi , Y_{n+1}\right]  - \frac{1}{2} 
\frac{\diff}{\diff t} \phi  \right)^\dagger \Psi_{n+1} - \left( \left[ \phi , 
Y_{n+1}\right]  - \frac{1}{2} \frac{\diff}{\diff t} \phi  
\right) \Psi_{n+1}^\dagger \Bigg\} \notag \\
&=\int_{\mathbb{R}} \diff t \ e^{2nt} \ \tr  \Bigg\{ \phi \cdot \im 
\Bigg[ \frac{\diff}{\diff t} (\Psi_{n+1} + 
\Psi_{n+1}^\dagger) + 2n (\Psi_{n+1} + \Psi_{n+1}^\dagger)  \\
&\qquad \qquad \qquad  + 2 \sum_{k=1}^{n+1}\left( \left[\Psi_k, 
Y_k^\dagger \right] + \left[Y_k, \Psi_k^\dagger \right] \right) \Bigg]
 \Bigg\} -\im \int_{\mathbb{R} } \frac{\diff}{\diff t} \left\{ e^{2nt} 
\tr \phi(\Psi_{n+1} + \Psi_{n+1}^\dagger)  \right\}  \; .
 \notag
\end{align}
\end{subequations}
A close inspection of the boundary term reveals that
\begin{align}
 \int_{\mathbb{R} } \frac{\diff}{\diff t} \left\{ e^{2nt} \ 
\tr \left( \phi(\Psi_{n+1} + \Psi_{n+1}^\dagger) \right) \right\}  =  e^{2nt} \
\tr \left( \phi(\Psi_{n+1} + \Psi_{n+1}^\dagger) \right) \Bigg|_{t\to 
-\infty}^{t\to 
+\infty} 
\end{align}
vanishes provided $\lim_{t\to\pm \infty}\phi(t)=0$ i.e. the map defined 
in~\eqref{eqn:moment_map} is a moment map for the action of the \emph{framed
gauge group} $\widehat{\Gcal}_0 = \{ g(t)| g: \R \to \urm(p) , 
\text{s.t. } \lim_{t\to \pm \infty}g(t) =1 \}$.
\subsection{Notation}
We need to introduce some notation, which is relevant for the proofs later.
\paragraph{$\partial$,$\bar{\partial}$-operators}
Following~\cite{Donaldson:1984}, we define the following $\partial$, 
$\bar{\partial}$-operators on $\C^p$-valued functions $f$ on $\R^-$
\begin{subequations}
\begin{alignat}{2}
 \diffZbar f &= \frac{1}{2} \frac{\diff}{\diff s} f + \Zcal f \;, & \qquad 
 \diffZ f &= \frac{1}{2} \frac{\diff}{\diff s} f - \Zcal^\dagger f \; ,\\
\diffYbar{j} f &= \Ycal_j f \; ,& \qquad  \diffY{j} f &= - \Ycal_j^\dagger f \; 
,
\end{alignat}
and on matrix-valued functions $\gamma$ on $\R^-$
\begin{alignat}{2}
 \diffZbar \gamma &= \frac{1}{2} \frac{\diff}{\diff s} \gamma + 
\left[\Zcal,\gamma \right] \; , & \qquad 
\diffZ \gamma &= \frac{1}{2} \frac{\diff}{\diff s} \gamma - 
\left[\Zcal^\dagger ,\gamma \right] \; ,\\
\diffYbar{j} \gamma  &= \left[ \Ycal_j , \gamma \right] \; , & \qquad
\diffY{j} \gamma &= - \left[\Ycal_j^\dagger, \gamma \right] \; .
\end{alignat}
\end{subequations}
These operators will give rise to the $\bar{\partial}$-operators associated to 
the connection $\Acal$. For that we take the covariant derivative $\diff_\Acal 
= \diff + \widehat{\Gamma}^P + Y_j \theta^j + Y_{\bar{j}}\bar{\theta}^j$ and 
define $\bar{\partial}_\Acal = \bar{\partial} + (\widehat{\Gamma}^P)^{(0,1)} +  
Y_{\bar{j}}\bar{\theta}^j$. Hence, the above definitions are understood as 
components of $\bar{\partial}_\Acal$. However, our notation and conventions 
differ slightly from~\cite{Donaldson:1984} in the sense that we work with the 
equivalent $\partial_\Acal$-operator. In detail, the cone direction $s$ 
in~\cite{Donaldson:1984} is considered as $0$th coordinate such that the 
\emph{canonical} complex structure is defined via the choice of $(1,0)$-forms 
$\diff s+\im e^1$ and $e^2+\im e^3$ ($\{e^p,\ p=1,2,3\}$ a co-frame on $\R^3$). 
In contrast, we designated the cone coordinate as $e^{2n+2}$ and choose the 
$(1,0)$-forms as in~\eqref{eqn:def_cplx_forms} in order to avoid unnecessary 
factors of $\im$. With respect to the canonical choice $e^{2j{-}1}+\im e^{2j}$ 
our complex structure is simply $J = -J_{\mathrm{can}}$, implying that we 
interchanged $(1,0)$ and $(0,1)$-forms. Consequently, we consider the 
$\partial_\Acal$-operator.
\paragraph{Gauge transformations}
For the $\partial$-operators the action of the complex automorphisms is 
defined via
\begin{align}
 \diffYbar{j}^g \coloneqq g \circ \diffYbar{j} \circ g^{-1} \and 
 \diffZbar^g \coloneqq g \circ \diffZbar \circ g^{-1} \; .
\end{align}
From these definitions, we obtain
\begin{subequations}
\begin{alignat}{2}
 g^{-1} \diffZbar^g g &= \diffZbar \; ,& \qquad g^{-1} \diffZ^g g &= \diffZ + 
h^{-1} \diffZ h \; ,\\
 g^{-1} \diffYbar{j}^g g &= \diffYbar{j} \; , & \qquad g^{-1} \diffY{j}^g g &= 
\diffY{j} +h^{-1} \diffY{j} h  \; .
\end{alignat}
\end{subequations}
for $h \coloneqq g^\dagger g$.
\paragraph{Complex equations}
The complex equations it holds
\begin{subequations}
\label{eqn:cplx_eqs_operators}
\begin{alignat}{3}
 \left[ \diffYbar{j},\diffYbar{k}\right] &=0 & \qquad &\Leftrightarrow & \qquad
\left[\Ycal_j,\Ycal_k \right]&=0 \; ,\\
\left[ \diffZbar,\diffYbar{j} \right] &=0 & &\Leftrightarrow  &
\frac{1}{2} \frac{\diff}{\diff s} \Ycal_j  &= \left[ \Ycal_j ,\Zcal \right] \; ,
\end{alignat}
\end{subequations}
where the right-hand-side is understood as acting on $\C^p$- or matrix-valued 
functions.

For the integrability of $\partial_\Acal$, i.e. $\partial_\Acal^2 =0$, we need 
besides~\eqref{eqn:cplx_eqs_operators} also $\partial_{\widehat{\Gamma}^P}^2 
=0$ and~\eqref{eqn:equivariance_cplx} to hold. Fortunately, 
$\widehat{\Gamma}^P$ is an HYM-instantons and, thus, defines an integrable 
$\partial$-operator. Moreover, by construction we restricted to matrix-valued 
fuctions 
$\Ycal_j$ and $\Zcal$ that satisfy the equivariance. In summary, the complex 
equations are the integrability conditions for $\partial_\Acal$.
\paragraph{Real equation}
Recall the definition~\eqref{eqn:moment_map} of the moment 
map $\mu(\Ycal,\Zcal)$. The expression is identical 
to the action of the operator~\footnote{This object is analogous to 
$\widehat{F}$ of~\cite[eq. (1.10)]{Donaldson:1984}.} 
\begin{equation}
  \Upsilon(\Ycal,\Zcal) \coloneqq 2 \left( \left[\diffZ,\diffZbar \right] + 
\lambda_n(s) 
\sum_{j=1}^n \left[ \diffY{j}, \diffYbar{j}\right] \right)
\end{equation}
in the usual sense.
This operator behaves under complex gauge transformations as follows
\begin{equation}
 g^{-1} \left( \Upsilon(\Ycal^g,\Zcal^g) \right) g = \Upsilon(\Ycal,\Zcal) 
- 2 \left( \diffZbar (h^{-1} \diffZ h)  + \lambda_n(s) \sum_{j=1}^n 
\diffYbar{j} (h^{-1} \diffY{j}h)\right) 
\; .
\end{equation}
\subsection{Adaptation of proofs}
\label{sec:proof}
\subsubsection{Differential inequality}
\label{subsec:diff_inequality}
Let $\{\kappa_i\}_{i=1,\ldots,p}$ be the \emph{positive} eigenvalues (still 
functions of $s$) of $h$ on $I_\epsilon$. Define
\begin{equation}
 \Phi(h) :=  \ln \left( \max_{i=1,\ldots,p} \kappa_i \right) \; ,
\end{equation}
which is well-defined. The claim is that the inequalities
\begin{subequations}
\label{eqn:diff_inequality}
\begin{align}
 \frac{\diff^2}{\diff s^2} \Phi(h) \geq -2 \left( \| \Upsilon(\Ycal,\Zcal) \| + 
\| \Upsilon(\Ycal^g,\Zcal^g) \| \right) \; ,\\
 \frac{\diff^2}{\diff s^2} \Phi(h^{-1}) \geq -2 \left( \| \Upsilon(\Ycal,\Zcal) 
\| + \| \Upsilon(\Ycal^g,\Zcal^g) \| \right)
\end{align}
\end{subequations}
hold in a weak sense.
\begin{Proof}
 Following~\cite{Donaldson:1984}, it is sufficient to consider the case where 
all eigenvalues of $h$ are distinct for each $s$.
Further, by a unitary gauge transformation one finds in each 
$\glrm(p,\C)\slash \urm(p)$-equivalence class an element $g$ (which 
corresponds to a given $h$) such that 
\begin{equation}
 g = \diag(e^{t_1} , \ldots, e^{t_p}) \with t_1(s) > t_2(s) > \ldots > t_p(s) 
\quad \forall s\in I_\epsilon \; .
\end{equation}
Hence, one obtains $h=\diag(e^{2t_1} , \ldots, e^{2t_p})$ and 
$h^{-1}=\diag(e^{-2t_1} , \ldots, e^{-2t_p})$ such that $\Phi(h)=2 t_1$ and 
$\Phi(h^{-1})=-2t_p$.
Next, we compute
\begin{subequations}
\begin{align}
 \diffZ h &= \diag(e^{2t_j} \tfrac{\diff}{\diff s} t_j) - \left[ 
\Zcal^\dagger,h \right] \; , \\
h^{-1}\diffZ h &=\diag( \tfrac{\diff}{\diff s} t_j ) + \Zcal^\dagger - h^{-1} 
\Zcal^\dagger h  \; ,\\
\diffZbar(h^{-1}\diffZ h) &= \diag\left( \frac{1}{2} \frac{\diff^2}{\diff s^2} 
t_j  \right) + \left[ \Zcal, \diag(\frac{\diff}{\diff s} t_j) \right] + 
\frac{1}{2} \frac{\diff}{\diff s} \left( \Zcal^\dagger - h^{-1} \Zcal^\dagger h 
\right) + \left[ \Zcal, \Zcal^\dagger - h^{-1} \Zcal^\dagger h \right] \; .
\end{align}
\end{subequations}
Now, we consider the diagonal elements
\begin{align}
 \left(\diffZbar(h^{-1}\diffZ h)\right)_{(a,a)} = \frac{1}{2} 
\frac{\diff^2}{\diff s^2} t_a + \sum_{b\neq a} |\Zcal_{ab}|^2 \left\{ 
\left(1-e^{2(t_a - t_b)}  \right) - \left( 1-e^{-2(t_a - t_b)} \right) \right\} 
\; ,
\end{align}
where we used
\begin{equation}
 \left( \left[ \Zcal , \diag(\frac{\diff}{\diff s} t_j) \right]\right)_{(a,a)} 
=0 \and \left( \Zcal^\dagger - h^{-1} \Zcal^\dagger h \right)_{(a,a)}=0 \; .
\end{equation}
Similarly, one derives
\begin{align}
 \left(\diffYbar{j}(h^{-1}\diffY{j} h)\right)_{(a,a)} = \sum_{b\neq a} 
|(\Ycal_j)_{ab}|^2 \left\{ \left(1-e^{2(t_a - t_b)}  \right) - \left( 
1-e^{-2(t_a - t_b)} \right) \right\} \; .
\end{align}
Then, one proceeds
\begin{align}
 \left( \Upsilon(\Ycal,\Zcal) -  \Upsilon(\Ycal^g,\Zcal^g) \right)_{(a,a)} 
&= 
 \left(\Upsilon(\Ycal,\Zcal) - g^{-1} \left( \Upsilon
(\Ycal^g,\Zcal^g)\right) g \right)_{(a,a)} \\
 &=  2 \left(  \diffZbar (h^{-1} \diffZ h)  + \lambda_n(s) \sum_{j=1}^n 
\diffYbar{j} (h^{-1} \diffY{j}h) \right)_{(a,a)} \notag \\
&= \frac{\diff^2}{\diff s^2} t_a + 2 \sum_{b\neq a} \left( |\Zcal_{ab}|^2 + 
\lambda_n(s) \sum_{j=1}^n |(\Ycal_j)_{ab}|^2 \right) \left\{ \left( 1-e^{2(t_a 
- t_b)} \right) - \left(1-e^{-2(t_a -t_b)} \right) \right\} \notag
\end{align}
To get the estimate for $\Phi(h) = 2 t_1$ take $a=1$ and use $\left\{ \left( 
1-e^{2(t_1 - t_b)} \right) - \left(1-e^{-2(t_1 -t_b)} \right) \right\} < 0$ as 
$t_1 > t_b$ for all $b>1$. Then
\begin{align}
 \frac{\diff^2}{\diff s^2} t_1 &\geq -\left( \Upsilon(\Ycal^g,\Zcal^g) 
- \Upsilon(\Ycal,\Zcal)   \right)_{(1,1)}
\geq -  \left( |\Upsilon(\Ycal^g,\Zcal^g)_{(1,1)}|+ 
|\Upsilon(\Ycal,\Zcal)_{(1,1)}|   
\right) \notag \\
&\geq -  \left( \|\Upsilon(\Ycal^g,\Zcal^g) \|+ 
\|\Upsilon(\Ycal,\Zcal)\|   \right) \notag  \\
&\Rightarrow \frac{\diff^2}{\diff s^2} \Phi(h) \geq - 2 \left( 
\|\Upsilon(\Ycal^g,\Zcal^g) \|+ \|\Upsilon(\Ycal,\Zcal)\|   \right)
\end{align}
Similarly, the estimate for $\Phi(h^{-1})$ is obtained by taking $a=p$ and 
$\left\{ \left( 1-e^{2(t_p - t_b)} \right) - \left(1-e^{-2(t_p -t_b)} \right) 
\right\} > 0$ for all $b<p$. Then
\begin{align}
 \left(\Upsilon(\Ycal,\Zcal) - \Upsilon(\Ycal^g,\Zcal^g) \right)_{(p,p)} 
\geq \frac{\diff^2}{\diff s^2} t_p \quad \Rightarrow \quad
\frac{\diff^2}{\diff s^2} \Phi(h^{-1}) \geq - 2 \left( 
\|\Upsilon(\Ycal^g,\Zcal^g) 
\|+ \|\Upsilon(\Ycal,\Zcal)\|   \right)
\end{align}
Thus, the claim~\eqref{eqn:diff_inequality} holds.
\end{Proof}
\subsubsection{Uniqueness}
\label{subsec:uniqueness}
Suppose that $(\Ycal,\Zcal)$ is a solution to the complex equations on 
$I_\epsilon$. Let us assume that we have two complex gauge 
transformations $g_1$ and $g_2$ such that 
\begin{enumerate}[(i)]
 \item $\mu(\Ycal^{g_1},\Zcal^{g_1})=0$ and $\mu(\Ycal^{g_2},\Zcal^{g_2})=0$ in 
$I_\epsilon$ 
\item $h_1 = g_1^\dagger g_1$ and $h_2=g_2^\dagger g_2$ satisfying $h_1 
|_{\partial I_\epsilon} = h_2 |_{\partial I_\epsilon}$.
\end{enumerate}
Then $h_1 = h_2$ in $I_\epsilon$.
\begin{Proof}
 We can suppose $g_2=1$ such that $h_2 =1$ in $I_\epsilon$ and $\partial 
I_\epsilon$. Hence, $g\equiv g_1$ and $h|_{\partial I_\epsilon}=1$. Since 
$\Upsilon(\Ycal,\Zcal)=0$ and $\Upsilon(\Ycal^{g},\Zcal^{g})=0$, we have 
\begin{align}
 \frac{\diff^2}{\diff s^2} \Phi(h) = 2\frac{\diff^2}{\diff s^2} t_1 \geq 0 
\text{ in } I_\epsilon  \; , \;  t_1|_{\partial I_\epsilon} = 0   \and  
\frac{\diff^2}{\diff s^2} 
\Phi(h^{-1}) = -2\frac{\diff^2}{\diff s^2} t_p \geq 0 \text{ in } I_\epsilon  
\; 
, \; t_p|_{\partial I_\epsilon} = 0 \; .
\end{align}
By (weak) convexity, it follows $t_1 \leq 0$ in $I_\epsilon$ and $t_p \geq 0$ 
in 
$I_\epsilon$, but we now arrive at $0 \geq t_1 > t_2 > \ldots > t_p \geq 0$. 
Hence, $t_j=0$ in $I_\epsilon$ and $h=1$ in $I_\epsilon$ (modulo unitary 
transformations).
\end{Proof}
\subsubsection{Boundedness}
\label{subsec:bounded}
Next, we need to show the boundedness of $\mu(\Ycal,\Zcal)$. The only critical 
term is $\lambda_n(s)$, which diverges for $s\to 0$. However, it is straight 
forward to derive the pole structure of the gauge transformed operator 
$\Upsilon$ to be
\begin{align}
\begin{split}
 g^{-1} \left( \Upsilon(\Ycal^g,\Zcal^g) \right) g \Big|_{\mathrm{pole}} &= 
\Upsilon(\Ycal,\Zcal)\Big|_{\mathrm{pole}} - 2 \lambda_n\sum_{j=1}^n 
\diffYbar{j} \left( h^{-1} \diffY{j} h \right) \\
&=  2 \lambda_n\sum_{j=1}^n \left[\Ycal_j , h^{-1} 
\Ycal_j^\dagger h \right]_{s\to0} \; .
\end{split}
\end{align}
But recall that we will consider framed gauge transformation, i.e. $h=1$ at the 
boundaries, and $\Ycal(s=0)$ are elements of a Cartan subalgebra. Hence, the 
potential pole vanishes for any gauge transformation once the correct boundary 
conditions~\eqref{eqn:boundary_cond} are imposed. Thus, $\mu(\Ycal,\Zcal)$ is 
bounded.
\subsubsection{Limit $\epsilon \to 0$}
\label{subsec:limit}
Finally, we need to show that the limit $\epsilon \to0$ exists, for which we 
follow~\cite{Kronheimer:1989,Kronheimer:1990}. Let $(\Ycal,\Zcal)$ be 
\emph{any} solution of the complex equation, then for each $\epsilon >0$ there 
exists a unique \emph{complex} gauge transformation $g_\epsilon$ such that 
$(\Ycal^{g_\epsilon},\Zcal^{g_\epsilon})$ satisfies the real equation in 
$I_\epsilon$. Associate $h_\epsilon = g_\epsilon^\dagger g_\epsilon$.

We start by constructing a solution $(\widehat{\Ycal} , \widehat{\Zcal})$ of 
the complex equations with the properties
\begin{align}
 (\widehat{\Ycal}_j , \widehat{\Zcal}) (s) = \Bigg\{\begin{matrix} ( \tau_j , 
0) & \for  & s=-\epsilon \\ (\T_j , \T_{n+1})& \for &-\tfrac{1}{\epsilon} < s 
<-1   \end{matrix} 
\end{align}
where $(\T_j,\T_{n+1})$ correspond to the complex linear combinations of the 
$T_\mu$ of the boundary condition~\eqref{eqn:boundary_cond}, i.e. they lie in 
a Cartan subalgebra of $\surmL(n+1)$. The $\tau_j$ are arbitrary points in the 
complex orbits 
$\mathcal{O}(\T_j)$, because we know that the boundary values at $s\to0$ are in 
gauge orbits of the $\T_j$.

The existence of such a solution follows from the local triviality of the 
complex equations. Note that this solution is \emph{constant} in 
$(-\tfrac{1}{\epsilon} ,-1  )$ and $\mu(\widehat{\Ycal} , \widehat{\Zcal}) 
=0$ for $-\tfrac{1}{\epsilon} < s <-1  $.

The claim then is: Starting from $(\widehat{\Ycal} , \widehat{\Zcal})$ as 
above, for each $\epsilon >0$ there exists a unique gauge transformation 
$g_\epsilon$ such that
\begin{enumerate}[(i)]
 \item $(\widehat{\Ycal}^{g_\epsilon} , 
\widehat{\Zcal}^{g_\epsilon})$ satisfies the real equation everywhere in 
$I_\epsilon$,
 \item $(\widehat{\Ycal}^{g_\epsilon} , \widehat{\Zcal}^{g_\epsilon})$ has the 
correct boundary conditions~\eqref{eqn:boundary_cond},
\item $g=1$ at the boundaries and $\widehat{\Zcal}^{g_\epsilon}$ is 
Hermitian,
\item $\Phi(h_\epsilon)$, $\Phi(h_\epsilon^{-1})$ are uniformly bounded.
\end{enumerate}
Thus, by the uniform bound, one has the existence a $C^\infty$ limit 
$h_{\infty} \coloneqq \lim_{\epsilon \to 0} h_\epsilon$ such that $g_\infty 
\coloneqq \sqrt{h_\infty}$ has all desired properties on the negative half-line.
\begin{Proof}
The existence and the uniqueness of such a $g_\epsilon$ follows from the 
above. Using the differential inequalities~\eqref{eqn:diff_inequality} and the 
boundedness of $\mu$ we derive at
\begin{align}
 \frac{\diff^2}{\diff s^2} \Phi(h_\epsilon) \geq \left\{ \begin{matrix} -2 
\|\Upsilon(\widehat{\Ycal} , \widehat{\Zcal}) \| \geq -2C & \for & -1 < s < 
-\epsilon \\
0 & \for &   -\tfrac{1}{\epsilon} < s <-1    \end{matrix} \right. \; .
\end{align}
Moreover, since $h_\epsilon=1$ at $\partial I_\epsilon$, the eigenvalues have 
to vanish, which implies $\Phi(h_\epsilon)=0=\Phi(h_\epsilon^{-1})$ at 
$\partial I_\epsilon$. Consider the bounded, continuous, non-negative function
\begin{align}
 f_\epsilon (s) &= \left\{ \begin{matrix}-C (s+1)(s+\epsilon) & \for & -1 < s < 
-\epsilon \\
0 & \for &   -\tfrac{1}{\epsilon} < s <-1   \end{matrix} \right.  \\
\with \frac{\diff^2 }{\diff s^2 } f_\epsilon  &= \left\{ \begin{matrix}-2C & 
\for & -1 < s < -\epsilon \\
0 & \for &   -\tfrac{1}{\epsilon} < s <-1   \end{matrix} \right.
\end{align}
in a weak sense. But then, we obtain
\begin{align}
 \frac{\diff^2}{\diff s^2} \left( \Phi(h_\epsilon) - f_\epsilon \right) \geq 0 
\text{ in } I_\epsilon \and \Phi(h_\epsilon) - f_\epsilon =0 \text{ at } 
\partial I_\epsilon \; .
\end{align}
By convexity, $\Phi(h_\epsilon) - f_\epsilon \leq 0$ in $I_\epsilon$, which 
then implies
\begin{align}
 \Phi(h_\epsilon) = 2 t_1 \leq  \left\{ \begin{matrix}-C (s+1)(s+\epsilon) \leq 
-C s (s+1) & \for & -1 < s < 
-\epsilon \\
0 & \for &   -\tfrac{1}{\epsilon} < s <-1   \end{matrix} \right. \; .
\end{align}
Applying the very same reasoning to $\Phi(h_\epsilon^{-1})$, we obtain 
$\Phi(h^{-1}) - f_\epsilon \leq 0$ in $I_\epsilon$ and thus
\begin{align}
 -\Phi(h_\epsilon^{-1}) = 2 t_p \geq   \left\{ \begin{matrix} C s (s+1) & \for 
& -1 < s < -\epsilon \\
0 & \for &   -\tfrac{1}{\epsilon} < s <-1   \end{matrix} \right. \; .
\end{align}
In conclusion, the eigenvalues of $h_\epsilon$ are uniformly bounded
\begin{equation}
 \frac{1}{2} f \geq t_1 > \ldots > t_p \geq - \frac{1}{2} f \for f(s) = 
\left\{ \begin{matrix}-C s(s+1) & \for & -1 < s < 
-\epsilon \\ 0 & \for &   -\tfrac{1}{\epsilon} < s <-1   \end{matrix} \right.
\end{equation}
independent of $\epsilon$. This uniform bound leads to the existence of the 
limit $\epsilon \to0$ of $h_\epsilon$.
\end{Proof}

%% file: Instanton_paper.bbl
\providecommand{\href}[2]{#2}\begingroup\raggedright\begin{thebibliography}{10}

\bibitem{Donaldson:1983}
S.~K. Donaldson, {\it {Self-dual connections and the topology of smooth
  {$4$}-manifolds}},  {\em Bull. Amer. Math. Soc. (N.S.)} {\bf 8} (1983), no.~1
  81--83.

\bibitem{Rajaraman:1982is}
R.~Rajaraman, {\em {Solitons and instantons. An introduction to solitons and
  instantons in quantum field theory}}.
\newblock North-Holland, Amsterdam, 1982.

\bibitem{Manton:2004tk}
N.~Manton and P.~Sutcliffe, {\em {Topological solitons}}.
\newblock Cambridge University Press, Cambridge, 2004.

\bibitem{Weinberg:2012}
E.~J. Weinberg, {\em {Classical solutions in quantum field theory}}.
\newblock Cambridge University Press, Cambridge, 2012.

\bibitem{Candelas:1985en}
P.~Candelas, G.~T. Horowitz, A.~Strominger, and E.~Witten, {\it {Vacuum
  Configurations for Superstrings}},  {\em Nucl.Phys.} {\bf B258} (1985)
  46--74.

\bibitem{Grana:2005jc}
M.~Grana, {\it {Flux compactifications in string theory: A Comprehensive
  review}},  {\em Phys.Rept.} {\bf 423} (2006) 91--158,
  [\href{http://arxiv.org/abs/hep-th/0509003}{{\tt hep-th/0509003}}].

\bibitem{Blumenhagen:2006}
R.~Blumenhagen, B.~Kors, D.~Lüst, and S.~Stieberger, {\it {Four-dimensional
  String Compactifications with D-Branes, Orientifolds and Fluxes}},  {\em
  Phys.Rept.} {\bf 445} (2007) 1--193,
  [\href{http://arxiv.org/abs/hep-th/0610327}{{\tt hep-th/0610327}}].

\bibitem{Gauntlett:2004yd}
J.~P. Gauntlett, D.~Martelli, J.~Sparks, and D.~Waldram, {\it {Sasaki-Einstein
  metrics on $S^2 \times S^3$}},  {\em Adv.Theor.Math.Phys.} {\bf 8} (2004)
  711--734, [\href{http://arxiv.org/abs/hep-th/0403002}{{\tt hep-th/0403002}}].

\bibitem{Gauntlett:2004hh}
J.~P. Gauntlett, D.~Martelli, J.~F. Sparks, and D.~Waldram, {\it {A New
  infinite class of Sasaki-Einstein manifolds}},  {\em Adv.Theor.Math.Phys.}
  {\bf 8} (2006) 987--1000, [\href{http://arxiv.org/abs/hep-th/0403038}{{\tt
  hep-th/0403038}}].

\bibitem{Cvetic:2005ft}
M.~Cvetic, H.~Lu, D.~N. Page, and C.~Pope, {\it {New Einstein-Sasaki spaces in
  five and higher dimensions}},  {\em Phys.Rev.Lett.} {\bf 95} (2005) 071101,
  [\href{http://arxiv.org/abs/hep-th/0504225}{{\tt hep-th/0504225}}].

\bibitem{Cvetic:2005vk}
M.~Cvetic, H.~Lu, D.~N. Page, and C.~Pope, {\it {New Einstein-Sasaki and
  Einstein spaces from Kerr-de Sitter}},  {\em JHEP} {\bf 0907} (2009) 082,
  [\href{http://arxiv.org/abs/hep-th/0505223}{{\tt hep-th/0505223}}].

\bibitem{Lu:2005sn}
H.~Lu, C.~Pope, and J.~F. Vazquez-Poritz, {\it {A New construction of
  Einstein-Sasaki metrics in $D\geq 7$}},  {\em Phys.Rev.} {\bf D75} (2007)
  026005, [\href{http://arxiv.org/abs/hep-th/0512306}{{\tt hep-th/0512306}}].

\bibitem{Harland:2009yu}
D.~Harland, T.~A. Ivanova, O.~Lechtenfeld, and A.~D. Popov, {\it {Yang-Mills
  flows on nearly Kähler manifolds and $G_2$-instantons}},  {\em
  Commun.Math.Phys.} {\bf 300} (2010) 185--204,
  [\href{http://arxiv.org/abs/0909.2730}{{\tt arXiv:0909.2730}}].

\bibitem{Harland:2010ix}
D.~Harland and A.~D. Popov, {\it {Yang-Mills fields in flux compactifications
  on homogeneous manifolds with SU(4)-structure}},  {\em JHEP} {\bf 1202}
  (2012) 107, [\href{http://arxiv.org/abs/1005.2837}{{\tt arXiv:1005.2837}}].

\bibitem{Bauer:2010fia}
I.~Bauer, T.~A. Ivanova, O.~Lechtenfeld, and F.~Lubbe, {\it {Yang-Mills
  instantons and dyons on homogeneous $G_2$-manifolds}},  {\em JHEP} {\bf 1010}
  (2010) 044, [\href{http://arxiv.org/abs/1006.2388}{{\tt arXiv:1006.2388}}].

\bibitem{Haupt:2011mg}
A.~S. Haupt, T.~A. Ivanova, O.~Lechtenfeld, and A.~D. Popov, {\it {Chern-Simons
  flows on Aloff-Wallach spaces and Spin(7)-instantons}},  {\em Phys.Rev.} {\bf
  D83} (2011) 105028, [\href{http://arxiv.org/abs/1104.5231}{{\tt
  arXiv:1104.5231}}].

\bibitem{Gemmer:2011cp}
K.-P. Gemmer, O.~Lechtenfeld, C.~Nölle, and A.~D. Popov, {\it {Yang-Mills
  instantons on cones and sine-cones over nearly Kahler manifolds}},  {\em
  JHEP} {\bf 1109} (2011) 103, [\href{http://arxiv.org/abs/1108.3951}{{\tt
  arXiv:1108.3951}}].

\bibitem{Harland:2011zs}
D.~Harland and C.~Nölle, {\it {Instantons and Killing spinors}},  {\em JHEP}
  {\bf 1203} (2012) 082, [\href{http://arxiv.org/abs/1109.3552}{{\tt
  arXiv:1109.3552}}].

\bibitem{Ivanova:2012vz}
T.~A. Ivanova and A.~D. Popov, {\it {Instantons on special holonomy
  manifolds}},  {\em Phys.Rev.} {\bf D85} (2012) 105012,
  [\href{http://arxiv.org/abs/1203.2657}{{\tt arXiv:1203.2657}}].

\bibitem{Bunk:2014kva}
S.~Bunk, T.~A. Ivanova, O.~Lechtenfeld, A.~D. Popov, and M.~Sperling, {\it
  {Instantons on sine-cones over Sasakian manifolds}},  {\em Phys.Rev.} {\bf
  D90} (2014), no.~6 065028, [\href{http://arxiv.org/abs/1407.2948}{{\tt
  arXiv:1407.2948}}].

\bibitem{Bunk:2014coa}
S.~Bunk, O.~Lechtenfeld, A.~D. Popov, and M.~Sperling, {\it {Instantons on
  conical half-flat 6-manifolds}},  {\em JHEP} {\bf 1501} (2015) 030,
  [\href{http://arxiv.org/abs/1409.0030}{{\tt arXiv:1409.0030}}].

\bibitem{Donaldson:1984}
S.~K. Donaldson, {\it {Nahm's equations and the classification of monopoles}},
  {\em Comm.Math.Phys.} {\bf 96} (1984), no.~3 387--407.

\bibitem{Kronheimer:1989}
P.~B. Kronheimer, {\it {A Hyper-Kählerian structure on coadjoint orbits of a
  semisimple complex group}},  {\em J. London Math. Soc.} {\bf 42} (1990)
  193--208.

\bibitem{Kronheimer:1990}
P.~B. Kronheimer, {\it {Instantons and the geometry of the nilpotent variety}},
   {\em J. Differential Geom.} {\bf 32} (1990), no.~2 473--490.

\bibitem{Correia:2009}
F.~P. Correia, {\it {Hermitian Yang-Mills instantons on Calabi-Yau cones}},
  {\em JHEP} {\bf 0912} (2009) 004, [\href{http://arxiv.org/abs/0910.1096}{{\tt
  arXiv:0910.1096}}].

\bibitem{Correia:2010}
F.~P. Correia, {\it {Hermitian Yang-Mills instantons on resolutions of
  Calabi-Yau cones}},  {\em JHEP} {\bf 1102} (2011) 054,
  [\href{http://arxiv.org/abs/1009.0526}{{\tt arXiv:1009.0526}}].

\bibitem{Nibbelink:2007}
S.~Groot~Nibbelink, M.~Trapletti, and M.~Walter, {\it {Resolutions of $C^n/Z_n$
  Orbifolds, their U(1) Bundles, and Applications to String Model Building}},
  {\em JHEP} {\bf 0703} (2007) 035,
  [\href{http://arxiv.org/abs/hep-th/0701227}{{\tt hep-th/0701227}}].

\bibitem{Boyer:2008}
C.~P. Boyer and K.~Galicki, {\em {Sasakian geometry}}.
\newblock Oxford University Press, Oxford, 2008.

\bibitem{Atiyah:1982}
M.~Atiyah and R.~Bott, {\it {The Yang-Mills equations over Riemann surfaces}},
  {\em Phil.Trans.Roy.Soc.Lond.} {\bf A308} (1982) 523--615.

\bibitem{Deser:2014}
A.~Deser, O.~Lechtenfeld, and A.~D. Popov, {\it {Sigma-model limit of
  Yang–Mills instantons in higher dimensions}},  {\em Nucl.Phys.} {\bf B894}
  (2015) 361--373, [\href{http://arxiv.org/abs/1412.4258}{{\tt
  arXiv:1412.4258}}].

\bibitem{Donaldson:1985}
S.~K. Donaldson, {\it {Anti self-dual Yang-Mills connections over complex
  algebraic surfaces and stable vector bundles}},  {\em Proc. London Math. Soc.
  (3)} {\bf 50} (1985), no.~1 1--26.

\bibitem{Uhlenbeck:1986}
K.~Uhlenbeck and S.~T. Yau, {\it {On the existence of hermitian-Yang-Mills
  connections in stable vector bundles}},  {\em {Commun. Pure Appl. Math.}}
  {\bf 39} (1986) S257--S293.

\bibitem{Donaldson:1987}
S.~K. Donaldson, {\it {Infinite determinants, stable bundles and curvature}},
  {\em Duke Math. J.} {\bf 54} (1987) 231--247.

\bibitem{Thomas:2006}
R.~P. Thomas, {\it {Notes on {GIT} and symplectic reduction for bundles and
  varieties}},  in {\em {Surveys in differential geometry}}, vol.~10 of {\em
  {Surv.Differ.Geom.}}, pp.~221--273.
\newblock Int. Press, Somerville, MA, 2006.

\bibitem{Donaldson:1992}
S.~Donaldson, {\it {Boundary value problems for Yang-Mills fields}},  {\em
  J.Geom.Phys.} {\bf 8} (1992) 89--122.

\bibitem{Popov:2010rf}
A.~D. Popov and R.~J. Szabo, {\it {Double quiver gauge theory and nearly
  Kähler flux compactifications}},  {\em JHEP} {\bf 1202} (2012) 033,
  [\href{http://arxiv.org/abs/1009.3208}{{\tt arXiv:1009.3208}}].

\bibitem{Lechtenfeld:2012yw}
O.~Lechtenfeld, {\it {Instantons and Chern-Simons flows in 6, 7 and 8
  dimensions}},  {\em Phys.Part.Nucl.} {\bf 43} (2012) 569--576,
  [\href{http://arxiv.org/abs/1201.6390}{{\tt arXiv:1201.6390}}].

\bibitem{Wolf:2012gz}
M.~Wolf, {\it {Contact Manifolds, Contact Instantons, and Twistor Geometry}},
  {\em JHEP} {\bf 07} (2012) 074, [\href{http://arxiv.org/abs/1203.3423}{{\tt
  arXiv:1203.3423}}].

\bibitem{Xu:2008}
F.~Xu, {\em {Geometry of $SU(3)$ manifolds}}.
\newblock PhD thesis, 2008.

\bibitem{Hitchin:1991}
N.~Hitchin, {\it {Hyper-Kähler Manifolds}},  {\em Asterisque} {\bf 206} (1992)
  137.

\bibitem{Lechtenfeld:2014fza}
O.~Lechtenfeld, A.~D. Popov, and R.~J. Szabo, {\it {Sasakian quiver gauge
  theories and instantons on Calabi-Yau cones}},
  \href{http://arxiv.org/abs/1412.4409}{{\tt arXiv:1412.4409}}.

\bibitem{Lechtenfeld:2015ona}
O.~Lechtenfeld, A.~D. Popov, M.~Sperling, and R.~J. Szabo, {\it {Sasakian
  quiver gauge theories and instantons on cones over lens 5-spaces}},  {\em
  Nucl. Phys.} {\bf B899} (2015) 848--903,
  [\href{http://arxiv.org/abs/1506.02786}{{\tt arXiv:1506.02786}}].

\end{thebibliography}\endgroup
